\documentclass[epj,nopacs]{svjour}

\usepackage{graphics}

\usepackage{amsmath}
\usepackage{graphicx}
\usepackage{multirow}
\usepackage[noautoscale,vcentermath]{youngtab}
\usepackage{amsmath,amstext,amsfonts,amsbsy,amssymb,amscd,bbm,epsfig,lscape,cite}

\newcommand{\be}{\begin{equation}}
\newcommand{\ee}{\end{equation}}
\newcommand{\bea}{\begin{eqnarray}}
\newcommand{\eea}{\end{eqnarray}}

\begin{document}

\title{Micro-orbits in a many-branes model and deviations from $1/r^2$ Newton's law}


\author{A.~Donini\inst{1} \and S. G. Marim\'on \inst{1} }
\institute{Instituto de F\'{\i}sica Corpuscular, CSIC-Universitat de Val\`encia,\\
       Apartado de Correos 22085, E-46071 Valencia, Spain
}

\date{Received: date / Revised version: date}

\abstract{
We consider a 5-dimensional model with geometry ${\cal M} = {\cal M}_4 \times {\cal S}_1$, with compactification radius $R$. 
The Standard Model particles are localized onto a brane located at y=0, with identical branes localized at different points in the extra dimension. 
Objects located on our brane can orbit around objects located on a brane at a distance $d=y/R$, with an orbit and a period significantly different from the standard Newtonian ones. We study the kinematical properties of the orbits, finding that it is possible to distinguish one motion from the other in a large region of the 
initial conditions parameter space. This is a warm-up to study if a  SM-like mass distribution on one (or more) distant brane(s) may represent a possible dark matter candidate. 
After using the same technique to the study of orbits of objects lying on the same brane ($d=0$), we apply this method
to detect generic deviations from the inverse-square Newton's law. We propose a possible experimental setup to look for departures from Newtonian motion 
in the micro-world, finding that an order of magnitude improvement on present bounds can be attained at the 95\% CL under reasonable assumptions.
}

\maketitle

\section{Introduction}
\label{sec:intro}

Even after the discovery of a scalar particle with a mass $m_H = 125.7 \pm 0.4$ GeV \cite{Agashe:2014kda} in 2012 by 
the ATLAS and CMS Collaborations (see Refs. \cite{Aad:2013wqa,Aad:2014aba,Aad:2014eva,Aad:2014eha}  and 
\cite{Chatrchyan:2013iaa,Chatrchyan:2013mxa,Khachatryan:2014ira} for recent results), 
it is well possible that the Standard Model be not the end of the story
for several theoretical and experimental reasons. First of all, the Standard Model cannot explain the observed {\em Dark Matter} component of the Universe energy density, 
$\Omega_{\rm DM} \sim 27$\%;  it  has no clue for the so-called {\em Dark Energy}  that should determine the observed accelerated expansion of the 
Universe, $\Omega_{\rm DE} \sim 0.68$\%;  the amount of CP violation in the Standard Model is not enough to explain Baryogenesis; and, eventually, 
the observation of non-vanishing neutrino masses cries for an extension of the Standard Model that could account for them (allowing, 
in some extensions, for a Baryogenesis-through-Leptogenesis scenario). In addition to these experimental hints, the Standard Model does not include gravity, 
for which a coherent (and unique) quantized theory is lacking. 
Most of the Standard Model extensions have been advanced to solve some of these problems, by considering it
as an effective low-energy theory that should be replaced by a more fundamental one at some scale such as, for example, the Planck scale, $M_P \sim 10^{19}$ GeV. 
Notice that $M_P$ is well above the electroweak symmetry breaking scale, $\Lambda_{\rm EW}
\sim 246$ GeV, though. This enormous spread sounded {\em unnatural}  \cite{'tHooft:1979bh} for long, originating the so-called {\em hierarchy problem}. Typical solutions, 
such as supersymmetry \cite{Dimopoulos:1981zb} or technicolor \cite{Weinberg:1975gm,Weinberg:1979bn,Susskind:1978ms}, assume that
new physics, responsible for the electroweak symmetry breaking, must be found not much above the electroweak scale. 
Both hypotheses, however, predict the existence of many new particles not seen up to now 
at the LHC. A different proposal to solve the hierarchy problem was advanced in the '90s \cite{Antoniadis:1990ew,Antoniadis:1997zg,ArkaniHamed:1998rs,Antoniadis:1998ig}: 
to explain the large hierarchy between $\Lambda_{\rm EW}$ and $M_P$ 
without introducing new physics in between, why don't we lower $M_P$, instead? This could be done assuming the existence of new spatial dimensions 
in excess of the observed three ones to which we are used to at human-being length scales. In order for these new 
dimensions to pass unnoticed to the eye of an observer, they must be compactified in such tiny volumes that direct observation through the measurement of
deviations from the inverse-square Newton's law for gravitational interactions is beyond the reach of current experiments \cite{Adelberger:2009zz}.  
If gravity may  propagate into the {\em bulk} $V_n \sim  (2 \pi R)^n$,  with $R$ a generic compactification radius (more complicated compactification schemes
may be envisaged), at very small distances compared with $R$ gravity would be $D$-dimensional
(where $D = 4 + n$, being $n$ the number of extra spatial dimensions) with a fundamental scale $M_D$. On the other hand, at distances much larger than $R$, 
gravity behaves as in 4-dimensions, with fundamental scale $M_P^2 \propto (2 \pi R)^n M_D^{n+2}$.
This relation between $M_P$ and the fundamental scale $M_D$ of a $D$-dimensional gravitational theory 
was first derived in Refs.~\cite{Antoniadis:1997zg,ArkaniHamed:1998rs} and \cite{ArkaniHamed:1998nn}.  The relation states that, if $V_n$ is large enough, 
the fundamental mass scale $M_D$ can be much lower than $M_P$ and, possibly, as low as the electro-weak symmetry breaking scale $\Lambda_{\rm EW}$, thus solving the hierarchy problem\footnote{Being a large compact volume the origin of a large 4-dimensional Planck mass, this solution to the hierarchy problem is called {\em Large Extra-Dimensions} (LED).}. For $n = 1$,  $R$ should be of astrophysical size to have $M_D \sim 1$ TeV. However, for $n \geq 2$ to lower $M_D$ down to some TeV's a sub-mm radius $R$ suffices, something that is not excluded by direct observation of deviations from the Newton's $1/r^2$ law:
present limits on new spatial dimensions gives $R \leq 44 \, \mu$m at 95\% CL for the largest extra-dimension compactified in a circle of radius $R$ \cite{Kapner:2006si}. 
A huge literature has been devoted to study the virtues and problems of LED models (see, for example, Ref.~\cite{Ponton:2012bi} and references therein), and experimental searches
at the LHC of signatures of extra-dimensions in high-energy particle scattering are ongoing (see Ref.~\cite{Agashe:2014kda} for a recent update on the LED searches status). 
Notice, however, that non-observation of the characteristic signatures of LED models at the LHC is pushing limits on $M_D$ well above the TeV scale, thus making them 
less appealing as an elegant solution to the hierarchy problem.

On the other hand, extra-dimensions may be motivated on their own as a possible framework for Dark Matter. In LED models, the Standard Model is added to gravity by 
introducing two separate terms in the action \cite{Sundrum:1998sj}, $S = S_{\rm gravity} + S_{\rm SM}$. 
Whilst $S_{\rm gravity}$ is the $D$-dimensional Einstein-Hilbert action, $S_{\rm SM}$ is the standard 4-dimensional action of the Standard Model. 
The SM fields are stuck onto a 4-dimensional surface called {\em brane}, a concept borrowed by string theory \cite{Polchinski:1996na}. 
Little has been said about an interesting possibility: if we may conceive a space-time in which Standard Model particles are bounded to live on a 4-dimensional surface embedded in a higher-dimensional bulk, what forbids the existence of other identical branes, with identical (or different) matter located on them? 
This hypothesis has not been studied in full  detail after having been advanced at the very beginning of the LED proposal at the end of the '90s 
(albeit, to our knowledge, not in scientific publications). In particular, little interest has been devoted to the possibility that SM-like matter located on a different brane at a distance $|\vec{y}| < 2 \pi |\vec{R}|$ from us in the extra-dimensions may represent a fraction (or the total) of the Dark Matter component in the Universe.
Notice that, for three-dimensional distances $r$ much larger than the compactification scale $R$, $r \gg R$, gravity behaves effectively as in 4-dimensions. Therefore, 
the extra matter located on different branes act identically to standard matter in our Universe, albeit only gravitationally, as gauge interactions are only allowed on directions longitudinal to the branes, and not transverse to them. The extra matter on other branes, therefore, behave exactly as Dark Matter (taking into account present
bounds on direct and indirect Dark Matter searches, from which only very tight upper bounds on non-gravitational cross-sections of Dark Matter particles with SM ones can be
derived, see for example Ref.~\cite{Queiroz:2016awc} for a recent review).

Several papers have dealt with isimilar ideas. For example, in Ref.~\cite{ArkaniHamed:1999zg}, the idea that the brane in which we live may be folded
many times in a small compact volume was pursued. If two foldings of the brane  happen to be very near at some point in the extra-dimensions, matter located
on them would interact gravitationally but not through gauge interactions (whose messengers should travel much longer than gravity), thus behaving as Dark Matter. 
The same would happen within the framework of what
is known as {\em mirror matter}: matter identical to SM matter, albeit forbidden to interact through gauge fields with SM particles because of a conserved parity number
(see, {\em e.g.}, Ref.~\cite{Foot:2001hc} and refs. therein). In both cases, SM-like matter can interact gravitationally with matter in our Universe but
not through other interactions. A lot of work has been devoted to these ideas, trying to fulfill all present cosmological and astrophysical bounds on the 
Dark Matter properties (see, for example, Refs.~\cite{Ignatiev:2003js,Ciarcelluti:2004ik,Ciarcelluti:2004ip,Ciarcelluti:2010zz} for the case of mirror matter).
One of the main problems for SM-like matter to represent the Dark Matter component of the Universe is the fact that data favours a non-dissipative, collisionless fluid
and not matter that, naively, would cluster and form structures identical to those present in the visible sky (see, for example, the literature on {\em Double Disk Dark Matter}
\cite{Fan:2013yva,McCullough:2013jma}). Attempts to make models of dissipative dark matter agree with observational data can be found, for example, in 
Refs.~\cite{Foot:2014uba,Foot:2014osa,Foot:2016wvj}

This paper, however, is not the place to perform a comprehensive study of a many-branes model with SM-matter located identically on two or more branes as a possible solution
to the Dark Matter abundance problem. We will leave this ambitious program, hopefully, to forthcoming publications.
We restrict ourselves to a more limited, albeit inspiring goal: to study the classical kinematical
behaviour of masses located on two distant (in the compact extra-dimension) branes under the effect of the $D$-dimensional gravitational field. 
We study the simplest case, one single extra spatial dimension compactified on a circle of radius $R$, whose size should be within the present bounds given above. 
For simplicity, we have fixed\footnote{We are aware that this model cannot solve the hierarchy problem (as, for a sub-mm size extra-dimension, 
$M_D \sim  5 \times 10^5$ TeV), that could however be solved adding more than one extra-dimension.} $R = 10$ $\mu$m.
 We have chosen the masses of a gravitational source $m$ on 
a distant brane ({\em there}) and of a test body $m^\prime$ on our brane ({\em here}) to values such that the typical three-dimensional distance $r$ varies 
in the range $r \in [1,100]$ $\mu$m, for which we expect to maximize the possible deviations from Newtonian dynamics.
We have then derived the range of angular velocities $\dot \theta$ for which the orbit of $m^\prime$ around 
the projection of $m$ on our brane, $\vec x_0$, are not open trajectories. For this choice of initial conditions, we expect from Newtonian gravity stable, periodic, 
elliptical orbits of $m^\prime$ around $\vec x_0$, being $\vec x_0$ one of the foci of the ellipse. On the contrary, we have found that the trajectory
 of $m^\prime$ around $\vec x_0$ in a two-branes 5-dimensional model may be either an open path or a bounded one, but cannot be a closed orbit.
Bounded orbits are generally not elliptical, not periodic and with revolution times that can change significantly from one revolution to the next.
A significant precession of the "periapsis" (defined as the point for which the distance between $m$ and $m^\prime$ is minimal) is also observed in the considered region of the initial conditions parameter space. 
In order to assess quantitatively for which particular initial conditions we could distinguish Newtonian dynamics from the two-branes 5-dimensional one, 
we have produced mock data describing some characteristics of the orbit in the latter model. For this study, we computed the distance at the periapsis and the "apoapsis" 
(the point for which the distance between $m$ and $m^\prime$ reaches a maximum) of $m^\prime$ from $\vec x_0$ and the time needed for $m^\prime$ to perform the first 
$2 \pi$-revolution around $\vec x_0$ (of course, a more complete study of the geometrical shape of the orbit on a time span larger than a single revolution may be done). 
We have then tried to fit the data using Newtonian dynamics (seeing if the orbit can be indeed described by an ellipse with a focus at $\vec x_0$ where a 4-dimensional
gravitational source of mass $M$, not necessarily identical to $m$, lies).
Our conclusion is that, in a {\em gedanken experiment} in which a mass $m^\prime$ is orbiting around "nothing" at $\vec x_0$
({\em i.e.} around the projection of $m$ on our brane), the measurement of a few of the geometrical and kinematical properties of the orbit is enough 
to distinguish the two models in a significant portion 
of the parameter space (depending, of course, on the distance $d = y/R$ of the two branes: the nearer, the more difficult the two models are to be distinguished).
We have found that the most important experimental information (apart from the observation of a precession of the periapsis) is the measurement of the time
needed to $m^\prime$ to perform a $2 \pi$-revolution around the projection of $m$. 
Of course, as long as only two objects are considered, the presence or not of other branes in the extra dimension in addition to the two branes where the two bodies
lie is irrelevant, as long as branes are transparent to gravity (see, however, Ref.~\cite{Carena:2002me}). However, 
our analysis could be straightforwardly extended to the case of several objects located on several different branes ({\em i.e.} in a truly {\em many-branes} model)
taking into account that the potential acting on $m^\prime$ is just the sum of the potentials originating from $n$ sources $m_i$ ($i = 1, \dots, n$) located at distance 
$\sqrt{ r_i^2 + y_i^2} $ from $m^\prime$.

Armed with the expertise acquired in the case in which $m$ and $m^\prime$ are located onto different branes, we have applied the same technique to the interesting
case $d = 0$, {\em i.e.} the case in which the two masses are on the same brane. In other words, may the measurement of the kinematical properties 
of the orbit of a mass $m^\prime$ around a gravitational field source $m$ in the micro-world be used to detect deviations from the $1/r^2$ Newton's law?
The answer, apparently, is yes. Consider a "planet" P of mass $m \sim 10^{-7}$ g and a "satellite" S with a mass $m^\prime \sim 10^{-9}$ g 
at a distance from P $r_0 = 190$ $\mu$m with an angular velocity $\dot \theta_0 = 1.8 \times 10^{-4}$ rad/s. 
The Newtonian orbit travelled by S around P has an apoapsis at the starting distance $r_0$ and a periapsis after half a revolution at a distance $\sim 40$ $\mu$m. 
The period of a $2 \pi$-revolution of S around P, with the initial conditions given above is $T_N \sim 7000$ s, {\em i.e.} approximately two hours! 
On the other hand, we have found that if the two masses are located onto a brane in a 5-dimensional space-time with an extra-dimension compactified
on a circle of radius $R = 10$ $\mu$m, the  distance of the periapsis can be less than a half with of the Newtonian one. When S approaches its periapsis, 
the gravitational field is much more intense than in the Newtonian case, and a gravitational slingshot effect is induced on S. For this reason, the orbit is completely
different: an almost elliptical orbit is followed by a very short and very fast nearly circular one. This pattern is repeated every time, with the major axis of the
almost elliptical section of the orbit precessing around P at the ratio of $\sim \pi/2$ every two revolutions. The time needed for S to orbit around P is non-constant: 
a revolution with $T_{\rm long} \sim T_N$ is followed by a second, very fast one, $T_{\rm short} \ll T_N$ (with $T_{\rm short}$ ranging between 100 s to 1000 s). 
Measuring several revolution times and fit them to a constant (as expected in  the Newtonian case) is, therefore, a very powerful tool to discriminate 
a gravitational potential different from the Newtonian one. 

Notice that, as both the source of the gravitational field $m$ and the test mass $m^\prime$ are on our brane, both can be manipulated. 
Therefore, we are no longer in the realm of a {\em gedanken experiment}. We have, therefore, applied the method outlined above 
to the case of a phenomenological modification of the Newtonian potential in the form of a Yukawa correction proportional to 
$\alpha \, G_N m m^\prime\, {\rm exp}(-r/\lambda)$, where $\alpha = 2 \cos d$ and $\lambda = R$ 
in the case of one compact extra-dimension (this way to parametrize deviations from the $1/r^2$ Newton's law is standard in the literature). 
A possible experimental setup that fulfills the basic requirements (even though it should be clearly studied further in all its details)
is the following: put a platinum planet P with mass $m_P \sim 10^{-7}$ g and radius $r_P = 10.3$ $\mu$m at the center of a 1 mm$^3$ laboratory in vacuum; 
introduce in the laboratory a diamagnetic satellite S with mass $m_S \sim 10^{-9}$ g (for a pyrolitic graphite sphere, $\rho = 2.2$ g/cm$^3$ and $r_S = 4.8$ $\mu$m);
insert the lab between two magnets with a magnetic field $B \sim 0.5$ T, such that the diamagnetic sphere may levitate to cancel the Earth gravitational field.
Once the diamagnetic sphere, at an initial distance from P $r_0 = 190$ $\mu$m is put into motion with an angular velocity $\dot \theta_0 = 1.8 \times 10^{4}$ rad/s
(for example by means of photo-irradiation), we can measure the times $T_n$ it takes to S to perform $n$ revolutions around P and compare with the constant Newtonian 
period $T_N$ expected for this particular choice of initial conditions. In this way, we have been able to derive the attainable exclusion limits at 95\% CL, 
finding that an upper limit of $\lambda < 2$ $\mu$m can be obtained for $\alpha = 2$ (to be compared with the 
present limit for one extra-dimension $R < 44$ $\mu$m at 95\% CL). Limits of a few microns can be put down to $\alpha \sim 10^{-3}$ (where for $\alpha > 10^5$ 
bounds below 1 $\mu$m can be obtained).
An important comment is that typical backgrounds that limit the sensitivity of experiments that test deviations from the $1/r^2$ law 
(such as Coulomb, dipolar or Van der Waals electrical forces) are irrelevant in this case as they correct the gravitational force with a $1/r^2$ dependence 
on the distance of S from P, and therefore, according to the Bertrand's theorem, may not induce precession of the orbit
(these backgrounds may only modify the constant revolution time $T_N$ and are, therefore, easily taken into account by looking for variations of the revolution time
along the orbit). Another important background, the Casimir force between the test sphere and the gravitational source, is negligible
as the test sphere is a diamagnetic object and not a conductive metal). We have checked also that general relativity corrections (that go with $1/r^4$ and may cause
a precession of the periapsis, as in the case of Mercury) are also negligible. In summary, our results are very promising and we plan to investigate further
the possibility to use kinematical measurements of orbits of micro-spheres at micro-distances to test the Newton's law. 

The paper is organized as follows:
in Sect.~\ref{sec:gravipotential} we remind the gravitational potential felt by a body of mass $m^\prime$ at a distance $d = y/R$ in the extra-dimension 
from the source $m$ of the gravitational field (as from Refs.~\cite{Kehagias:1999my,Floratos:1999bv}); in Sect.~\ref{sec:graviforce} we compute the
gravitational force acting on $m^\prime$ in the case when $m^\prime$ is located on a brane at a distance $d$ in the extra-dimension from the source
 (this was first done in Ref.~\cite{Liu:2003rq}); in Sect.~\ref{sec:examples}
we study the motion of $m^\prime$ under the effect of the gravitational field induced by $m$ when the two bodies are on distant branes for masses, 
distances and angular velocities such that orbits range from tens to hundreds of microns and quantify statistically the region of the initial conditions parameter
space for which the orbit can be distinguished from a Newtonian one; in Sect.~\ref{sec:d0} we apply the same technique to the case when $m$ and $m^\prime$
lie on the same brane; in Sect.~\ref{sec:experiment} we extend our analysis to the study of general deviations from the $1/r^2$ Newton's law using the
kinematical properties of micro-orbits; eventually, in Sect.~\ref{sec:concl} we draw our conclusions.

\section{Gravitational potential in ${\cal M}_4 \times {\cal S}_1$}
\label{sec:gravipotential}

When the original Large Extra-Dimensions model was presented in Refs.~\cite{ArkaniHamed:1998rs,Antoniadis:1998ig}, a simple phenomenological potential 
was derived in the limit of very large standard dimensions $r = |\vec{r}|$ with respect to the average compactification radius $R = |\vec R|$, 
\begin{equation}
V_{4+n} (|\vec r| \gg |\vec R|) \sim - \frac{m \, m^\prime }{M^{2 + n}_D R^n r} \sim - \frac{m \, m^\prime}{ M^2_P r} \, ,
\end{equation}
where $m$ is the source of the gravitational field, $m^\prime$ a test mass and $M_P$ and $M_D$ are the Planck mass and the fundamental scale of gravity in $D = 4+ n$ dimensions, respectively. The last equation establishes a relation between the two scales: 
\begin{equation}
M_P^2  \sim M_D^{2+n} R^n \, ,
\end{equation}
so that the Planck scale can be much higher than the fundamental scale of gravity $M_D$ if the compact volume $V_n \propto R^n$ is large, 
thus solving the hierarchy problem. In a subsequent paper, Ref.~\cite{ArkaniHamed:1998nn}, the size of the first order corrections in $|\vec{r}|/R$ was also sketched: 
\begin{equation}
V_{4+n} (|\vec r| \gg |\vec R|) \simeq - \frac{m \, m^\prime}{M^2_P} \sum_{(k_1, \dots,k_n)} \frac{e^{- 2 \pi L |\vec k|/r}}{r} \, .
\end{equation}
A complete computation of the gravitational potential in the case of ${\cal M}_4 \times {\cal S}_n$, however, was only given in Refs.~\cite{Kehagias:1999my,Floratos:1999bv}. A very simple derivation of the potential can be found in Ref.~\cite{Liu:2003rq} and it is outlined below for the case at hand 
of one compact extra-dimension, only.

Consider, first, the gravitational potential generated by the mass $m$ in 5 non-compact dimensions acting on the test mass, $m^\prime$:
\begin{equation}
\label{eq:noncompactpotential5d}
V^{\rm non-compact}_5 (r, y) = - \frac{G_5 m \, m^\prime }{2} \frac{1}{\left[r^2 + y^2 \right]} \, ,
\end{equation}
where $l_0= \sqrt{ r^2 + y^2}$ is the distance from the source of the potential, divided into its three-dimensional projection $r = |\vec r|$ and 
its extra-dimensional component, $y$. The 5-dimensional Newton constant, $G_5$ is defined as $G_5 \equiv M_D^{-3}$, being $M_D$ the fundamental 
scale of gravity. 

Notice, however, that if we consider now an extra-dimension compactified on a circle of radius $R$, the path of length $l_0$ is not the only one 
that connects the mass $m^\prime$ with $m$: we can reach the source of the potential by traveling along a straight line {\em wrapping} 
around the compact dimension as many times as we want. The length of a path that goes $k$ times around the compact dimension is 
$l_k = \sqrt{|\vec r|^2 + (y - 2 \pi R k)^2}$. Therefore, the source is effectively {\em felt} by the mass $m^\prime$ infinitely many times, albeit the gravitational potential 
is increasingly feebler as long as we turn more and more. In order to compute the full gravitational potential felt by $m^\prime$ in a compact space-time, 
we can imagine an infinite extra-dimension $y$ with an infinite number of sources $m$ located at distance $2 \pi R$ from each other, and just sum their potentials:
\begin{equation}
V^{\rm compact}_5 (r, y) = - \frac{G_5 m \, m^\prime }{2} \sum_{k = -\infty}^\infty \frac{1}{\left[r^2 + (y - 2 \pi R k)^2 \right]} \, ,
\end{equation}
where the sum goes from $-\infty$ to $+\infty$ since we can wrap around the compact dimension traveling in both directions. Define $L = 2 \pi R$ the length
of the compact dimension. Then, use the following identity:
\begin{equation}
\frac{1}{r^2 + (y - L k)^2} = \frac{1}{2 i L r} \left ( \frac{1}{k + z} - \frac{1}{k + z^\star}  \right )   \, ,
\end{equation}
where
\begin{equation}
z = - \frac{y + i r}{L} \, .
\end{equation}

\begin{figure*}[ht]
\centering
\begin{tabular}{cc}
\includegraphics[width=7cm]{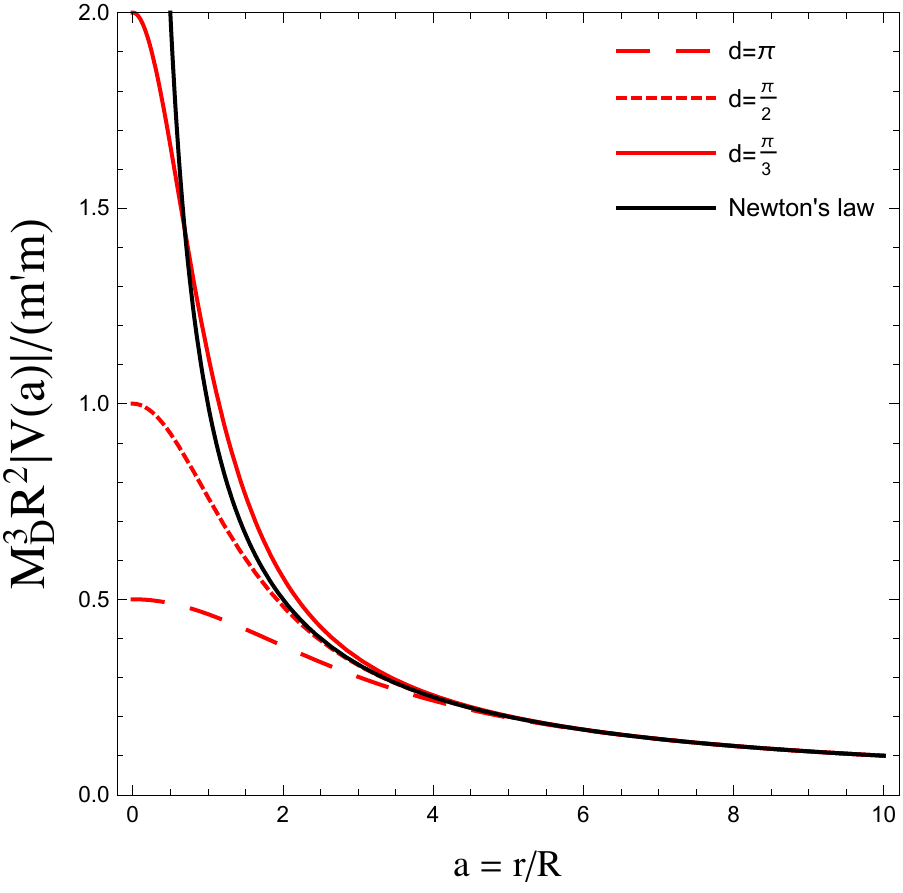} &
\includegraphics[width=7cm]{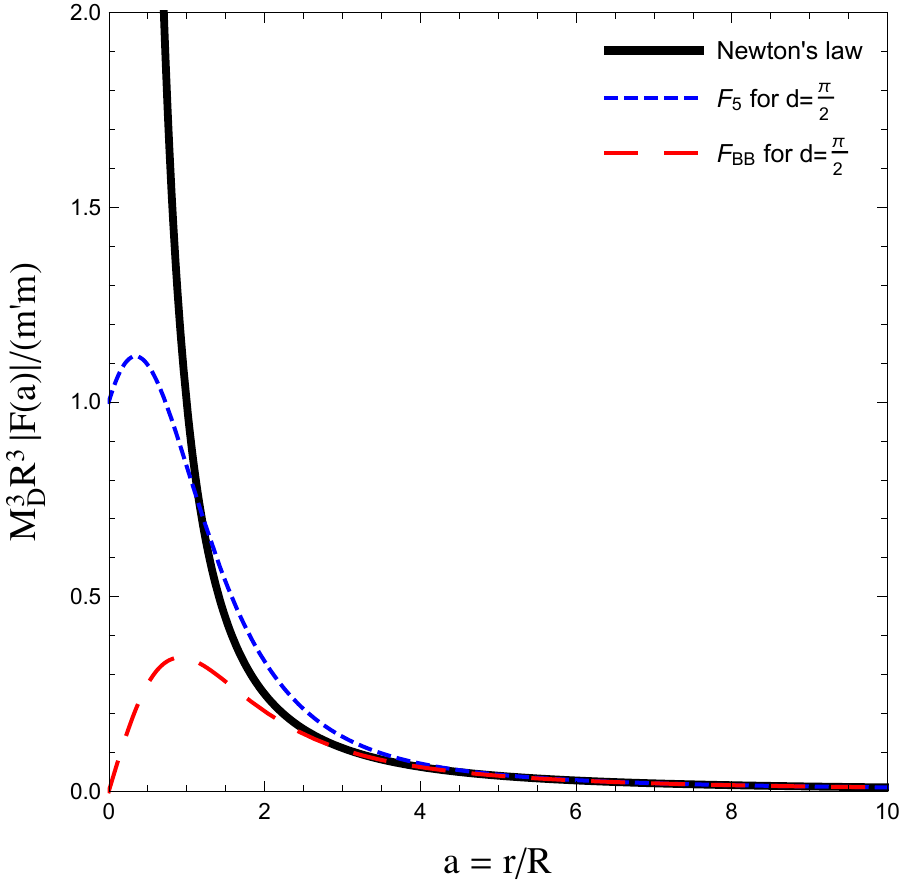}
\end{tabular}
\caption{\it 
Left panel: The dependence of the 5-dimensional potential $V_5^{\rm compact}(r,y)$ on the three-dimensional distance $a = r/R$ for three 
different values of $d = y/R$, $d  = \pi/3$ (red dashed line), $d=\pi/2$ (red dotted line) and $d=\pi$ (red solid line), for $R=10$ $\mu$ m.The Newton potential is plotted with a black solid line as a reference.
Right panel: Comparison of the different forces: the 4-dimensional Newton force (black solid line), the 5-dimensional gravitational force $|\vec{F}_5 (a,d)|$ (blue dashed line) and 
the brane-to-brane force $|{\vec F}_{\rm BB} (a,d)|$ (red dotted line), for $d = \pi/2$. All forces are properly rescaled in the vertical axis so as to be comparable in adimensional units.}
\label{fig:5dpotential}
\end{figure*}

The potential can thus be written as:
\begin{equation}
V_5^{\rm compact} (r, y) = - \frac{G_5 m \, m^\prime}{4 i L r} \sum_{k = -\infty}^\infty \left (  \frac{1}{k + z} - \frac{1}{k + z^\star} \right ) \, ,
\end{equation}
an expression that can be easily summed since: 
\begin{equation}
\sum_{k= -\infty}^\infty \frac{1}{k + z} = \pi \cot \pi z \, ,
\end{equation}
and, therefore, 
\begin{equation}
V_5^{\rm compact} (r, y) = - \frac{G_5 m \, m^\prime }{8 i R r}  \left ( \cot \pi z - \cot \pi z^\star \right ) \, .
\end{equation}
 After some algebraic manipulation, we get: 
 \begin{equation}
\label{eq:PotentialLiu5d}
V_5^{\rm compact} (r, y) = - \frac{G_5 m \, m^\prime}{4 R r} \left [ \frac{\sinh\left ( \frac{r}{R}\right ) }{\cosh \left ( \frac{r}{R}\right ) - \cos \left ( \frac{y}{R} \right ) } \right ] \, .
 \end{equation}
The 5-dimensional potential $V_5^{\rm compact}(r,y)$ as a function of the normalized three-dimensional distance $a = r/R$ is shown in Fig.~\ref{fig:5dpotential}(left) for three different values of the normalized distance in the bulk $d = y/R$: $d = \pi/3, \pi/2$ and $\pi$ (light solid, dotted and dashed lines, respectively). 
As it can be clearly seen, for $a \gg 1$ the potential does not depend on $d$ and becomes identical to the Newtonian 4-dimensional potential (depicted as
a bold solid line).
On the other hand, when $a \sim 1$, the distance $d$ plays a major role in determining 
the strength of the potential. A very important point to stress is that, for $y \neq 0$, there is no divergence at $r \to 0$, as the test mass at $\vec l = (\vec r, y)$ 
is not (yet) falling into the potential well located at $\vec l \to \vec 0$ but it remains at a safe distance $y$ from it.

The limits of small and large $a$ can be easily computed, albeit making a distinction between the case $y = 0$ and $y \neq 0$. 
For two masses located  on the same brane, $y = 0$, at very short three-dimensional spatial distance from the source we get:
 \begin{equation}
 \label{eq:Liupotentialall1yeq0}
V_5^{\rm compact} (a \ll 1, 0) \sim - \frac{G_5 m \, m^\prime }{2 R^2 a^2} + {\cal O} (a) \, ,
 \end{equation}
 {\em i.e.} the non-compact 5-dimensional potential of eq.~(\ref{eq:noncompactpotential5d}). 
 On the other hand, when $y \neq 0$, the potential is quite different: 
 \begin{equation}
\label{eq:Liupotentialall1yneq0}
V_5^{\rm compact} (a \ll 1, d) \sim - \frac{G_5 m \, m^\prime}{4 R^2 (1-\cos d)} + {\cal O} (a) \, ,
 \end{equation}
 as it is dominated by a volume term depending on the size of the extra-dimension. Notice that, since the gravitational force attracts necessarily a body in the 
 bulk towards the source of the potential, considered fixed onto a brane, at some time eq.~(\ref{eq:noncompactpotential5d}) must be recovered.
 
When the projection of the vector $\vec l$ onto the standard three spatial dimensions $r$ is much larger than the compactification radius $R$, $a \gg 1$, we have:
 \begin{equation}
\label{eq:Yukawapotential}
V_5^{\rm compact} (a \gg 1, d) \sim- \frac{G_5 m \, m^\prime}{4 R^2 a} \left [ 1 + 2 \cos d \,  e^{-a} + \dots
\right ] \, .
 \end{equation}
The leading term of eq.~(\ref{eq:Yukawapotential}) is nothing but the standard Newtonian 4-dimensional potential, after identifying:
\begin{equation}
\label{eq:5dimNewtonconstant}
G_4 \equiv \frac{G_5}{4 R} \, .
\end{equation}
The leading correction, on the other hand,  introduces a Yukawa-like potential whose impact can be experimentally tested (see Refs.~\cite{Adelberger:2009zz,Kapner:2006si}). 

\section{Gravitational force in ${\cal M}_4 \times {\cal S}_1$}
\label{sec:graviforce}

From the potential $V_5^{\rm compact} (r,y)$ it can be easily derived the gravitational force acting on a body of mass $m^\prime$ located in the 
bulk at distance $l_0 = \sqrt{r^2+ y^2}$ from the source of the gravitational field. 
We have: 
\begin{eqnarray}
\label{eq:force5dbranetobulk}
\frac{{\vec F}_5}{m^\prime} &=& - \frac{1}{m^\prime} \, \vec \nabla \, V_5 \nonumber \\
&=& - G_5 m \sum_{k = -\infty}^{ \infty}\frac{1}{\left[r^2+(y-2\pi Rk)^2\right]^{3/2}}\, \widehat{l_k} \nonumber \\
&=& - G_5 m \sum_{k=-\infty}^{ \infty}\frac{1}{l_k^3}\, \widehat{l_k} \ ,
\end{eqnarray}
where $l_k = \sqrt{r^2 + (y - 2 \pi R k)^2}$ and $\widehat l_k$ is a unit vector pointing in the direction of the mass $m^\prime$ from the source 
(that depends on the winding number $k$). 

The gravitational force that acts on a mass $m^\prime$ in the bulk under the effect of a mass $m$ located on a brane has been also computed
in Refs.~\cite{Kehagias:1999my,Floratos:1999bv}. An interesting consequence of eq.~(\ref{eq:force5dbranetobulk}) is that, given enough time, 
any mass located in the bulk will eventually be attracted towards the mass distribution located on the brane and, therefore, the bulk is necessarily
empty. The brane acts, in practice, as a "bulk vacuum-cleaner". On the other hand, this is not true if a mass is stuck
to a second brane, different from the one onto which is located the source of the gravitational field. This case has not been treated in the references above, 
but it has been studied in Ref.~\cite{Liu:2003rq}, instead. 

Consider the mass $m^\prime$ at a distance $l_0 = \sqrt{r^2 + y^2}$ where $y$ is the distance along the fifth-dimension between two parallel branes. 
Since $m^\prime$ cannot escape its own brane, the gravitational force originating at the location of $m$ is partially cancelled. The problem resembles, 
therefore, that of a mass onto an inclined plane, for which only the component of the force that goes along the plane remains. 
To compute the component of the brane-to-brane force along the second brane, we must derive the potential along $\vec r$: 
\begin{eqnarray}
\label{eq:force5dbranetobrane}
\frac{{\vec F}_{\rm BB}}{m^\prime} &=& - \left . \frac{1}{m^\prime} \, \vec \nabla \right |_{\vec r} \, V_5 \nonumber \\
&=& - G_5 m \sum_{k = -\infty}^{ \infty}\frac{\cos \theta_k}{\left[r^2+(y-2\pi Rk)^2\right]^{3/2}}\, \widehat{r} \nonumber \\
&=& - G_5 m \sum_{k=-\infty}^{ \infty}\frac{r}{l_k^4}\, \widehat{r} \ ,
\end{eqnarray}
with $\theta_k$ the angle between the vector $\vec l_k$ and our brane, and $\widehat r$ the (unique) unit vector along the projection of $\widehat l_k$ onto our brane.
Introducing the normalized coordinates $a = r/R$ and $d = y/R$ we get: 
\begin{equation}
\label{eq:FBB}
{\vec F}_{\rm BB} =  -  \frac{G_5 m \, m^\prime}{4 R^3 a^2} \, f_{\rm BB} (a,d)  \, \widehat r  \, ,
\end{equation}
where
\begin{equation}
\label{eq:forceddependence}
f_{\rm BB} (a,d) =  \left[ \frac{\sinh a }{\left ( \cosh a - \cos d \right )} -a\, \frac{1-\cosh  a \; \cos d}{(\cosh a-\cos d)^2} \right] \, .
\end{equation}

Notice that ${\vec F}_{\rm BB}$ is quite different from the well-known 4-dimensional Newton force: first of all, it is singular at $a \to 0$ only for $d= 0$, {\em i.e.} when the two masses
are on the same brane; on the other hand, for $d \neq 0$, the force vanishes as $a$ goes to zero, since the gravitational attraction felt by $m^\prime$ under the effect of $m$ 
cancels exactly with the constraint that bounds $m^\prime$ to remain on a brane at distance $d$ from the source.
The behavior of $|{\vec F}_{\rm BB}|$ as a function of $a$ is shown in Fig.~\ref{fig:5dpotential}(right): the black (solid) line represents the 4-dimensional Newton force, 
to be compared with the blue (dashed) line that represents the 5-dimensional force $|\vec{F}_5 (a,d)|$ acting on a particle at a distance $l_0 = R \sqrt{a^2 + d^2}$ from the source 
for the particular case $d = \pi/2$. 
On the other hand, the red (dotted) line represents the brane-to-brane force $|{\vec F}_{\rm BB} (a,d)|$ computed in eq.~(\ref{eq:FBB}) acting on a particle at a distance $l_0$ from the source but bounded to a second brane at a distance $d$ from our brane. First of all notice that both $|\vec{F}_5 (a,d)|$  and $|{\vec F}_{\rm BB} (a,d)|$ coincides with the 4-dimensional Newton force for $a \geq 4$ ({\em i.e.} above the present experimental bound on $R$, as they should). In the region $a \in [1,4]$ the 5-dimensional force $|\vec{F}_5 (a,d)|$ is larger than 
the 4-dimensional Newton force, contrary to the naive expectation that is deduced by applying the Gauss theorem to a non-compact space-time. For $a <1$ the 4-dimensional 
Newton force eventually becomes larger than its 5-dimensional counterpart, diverging for $a \to 0$ (whereas $|\vec{F}_5 (a \to 0 ,d)|$ goes to a constant). 
The brane-to-brane force $|{\vec F}_{\rm BB} (a,d)|$ is almost identical to the Newton force for $a \geq 2$, whereas the effect of both compactification and of the second-brane constrain becomes dominant for $a < 1$, eventually making $|{\vec F}_{\rm BB} (a,d)|$ vanish for $a \to 0$. Eventually, 
notice that both the brane-to-brane and the 5-dimensional force have a maximum for $a \sim 1$. 

The small $a$ limit of the brane-to-brane force is: 
\begin{equation}
|{\vec F}_{\rm BB} (a \ll 1, d) \simeq - \frac{G_5 m \, m^\prime}{12 R^3} \, a \, \frac{(2 + \cos d)}{(1- \cos d)^2} + {\cal O} (a^3) \, .
\end{equation}
On the other hand, for $a \gg 1$ we have: 
\begin{equation}
\label{eq:BBforceagg1}
|{\vec F}_{\rm BB} (a \gg 1, d) \simeq - \frac{G_5 m \, m^\prime}{4 R^3} \left [ \frac{1}{a^2} + 2 \cos d \frac{e^{-a}}{a}  + \dots \right ] \, ,
\end{equation}
where the first term in the expansion gives the $1/r^2$ 4-dimensional Newton's law. Notice that, depending on $d$ $F_{\rm BB}$ may be smaller or
larger than the Newtonian 4-dimensional force. 

Using eq.~(\ref{eq:BBforceagg1}), an upper bound on the compactification radius has been derived, $R \leq 44$ $\mu$m \cite{Agashe:2014kda}. The lower bound on the fundamental mass scale $M_D$ can then be derived using eq.~(\ref{eq:5dimNewtonconstant}):  we get $M_D \geq 5.5 \times 10^5$ TeV (well beyond LHC reach). 
Notice that, even if t$M_D$ is much lower than the Planck scale $M_P$, adding only one extra spatial dimension is not enough
to solve the hierarchy problem and bring the fundamental scale of gravity down to the electroweak scale as a huge hierarchy between $M_D$ and $\Lambda_{\rm EW}$ still exists.
On the other hand, for two extra spatial dimensions (for which the experimental bound on $R$ gives $R \leq 37 $ $\mu$m), the lower bound on $M_D$ becomes 
$M_D \geq 3.6$ TeV, within the reach of LHC. Recent limits put by both ATLAS and CMS using different signals imply that $M_D$ should be greater than a few TeV 
(see Ref.~\cite{Agashe:2014kda} and updates).

\section{Two bodies on different branes: a {\em gedanken} experiment}
\label{sec:examples}

Consider now two bodies located on two different branes at a distance $d = y/R$ in the extra dimension, with $R$ fixed to a value allowed by the present bound, 
$R = 10$ $\mu$m (we have checked that our results do not change  significantly for $R \in [10,50]$ $\mu$m, after proper tuning of the initial conditions). 
For simplicity, we fix the source mass $m$ on a distant brane ({\em i.e. there}) and the test mass $m^\prime$ onto our brane ({\em i.e. here}).  
As a consequence, we cannot interact with the source of the gravitational potential (that is out of our experimental reach), whereas
we can manipulate the test mass $m^\prime$: for example, we can choose its mass, its position and its velocity. 
The question we want to address is the following: can we distinguish the motion of $m^\prime$ induced by $m$ from a 4-dimensional Newtonian motion? 
Clearly, this experiment is not feasible in practice, as we have no handle to control the source, and for this reason it is a {\em gedanken} experiment.
What we can learn from it, however, is interesting in itself, as we will see that just by simple classical measurements of the geometry and period of the
motion of $m^\prime$ onto our brane under the effect of the gravitational force induced by an unseen source is enough to exclude a $1/r^2$ Newtonian force as the 
cause of such a motion. 

As a warm up, we first consider the case of a linear motion in Sect.~\ref{sec:linear}. Eventually, we study the two-dimensional case in Sect.~\ref{sec:orbit}.

\subsection {Linear motion}
\label{sec:linear}

Consider the mass $m$ in a brane at distance $d = y/R$ in the bulk. The projection of its position onto our brane, ${\vec x}_0$,
is taken to be the origin of a three-dimensional coordinate system , ${\vec x}_0 = {\vec 0}$.  The test mass $m^\prime$ is located onto our brane at a position $\vec x$, 
such that the distance in three dimensions  between the two masses is $r = |{\vec x}- {\vec x}_0 |$.
If we take the mass $m^\prime$ to be at rest or with an initial velocity aligned with the attracting gravitational force ${\vec F}_{\rm BB} (r, d)$, 
the resulting motion will be a linear motion. As there is no massive body located at ${\vec x}_0$ (the source is displaced at a distance $d$ in the extra-dimension), 
the test mass $m^\prime$ will not crash onto $m$. Quite the contrary, it will proceed in its motion, escaping from the source $m^\prime$ or being bounded in a periodic
motion in proximity of ${\vec x}_0$ depending on the initial conditions. 

Reducing the problem to a one-dimensional motion along the line that goes from ${\vec x}$ to ${\vec x}_0$, we must solve:
\begin{equation}
\ddot{r} = \frac{F(r) }{m^\prime} \, ,
\end{equation}
where $F (r)$ is either $F_4 (r)$ in the case of a 4-dimensional Newton force or $F_{\rm BB} (r, d)$ in the case of a brane-to-brane force between particles on branes at distance $d$
in the extra-dimension. In the first case,  we have:
\begin{equation}
\ddot{r}(t) + \frac{G_4 m}{r^2(t)}=0 \, .
\end{equation}
For simplicity, we will consider the mass $m^\prime$ small enough to neglect the motion of $m$ under the effect of $m^\prime$. 
Let's normalize the distance between the two bodies to the compactification radius $R$ introducing the normalized distance $a = r/R$. 
The differential equation to be solved is, thus: 
\begin{equation}
\ddot{a}(t) + \frac{k}{a^2 (t)} = 0 \, ,
\end{equation}
where $k = G_4 m / R^3$ is a coefficient with dimensions $s^{-2}$.  Since $R$ is bounded to be below 44 $\mu$m, the 
distance at which we want to compare the 4-dimensional Newtonian motion with the brane-to-brane case is $ r \in [1,100]$ $\mu$m. 
If we choose a mass $m \sim 10^{-7}$ g, then $G_4 m \sim 1$ $\mu$m$^3$/s$^2$ ({\em i.e.} $k \sim 10^{-3}$ s$^{-2}$) 
and $a$ is naturally of the required order. 

If the two bodies are on different branes, we have: 
\begin{equation}
\ddot{a (t)} + \frac{k_5 }{a^2 (t)} \, f_{\rm BB} (a,d) =  0 \, ,
\end{equation}
where $k_5 = G_5 m / 4 R^4$ (using the asymptotic relation in eq.~(\ref{eq:5dimNewtonconstant}) we have, trivially, $k_5 = G_4 m / R^3 = k$). 

\begin{figure*}[ht]
\centering
\includegraphics[angle=0,width=0.9\textwidth]{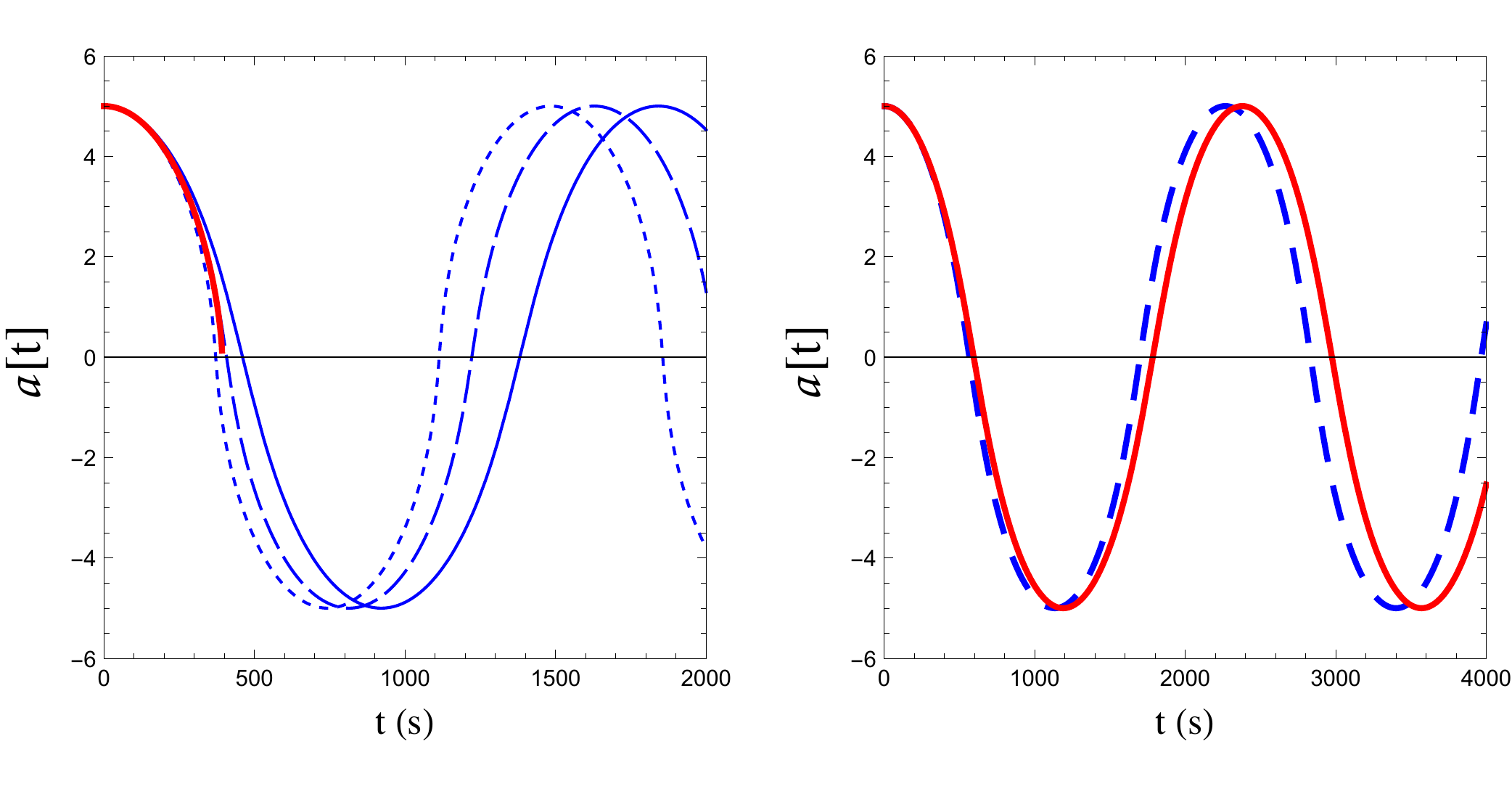}
\caption{\it Normalized distance $a$ between a mass $m^\prime$ and the source of a gravitational field $m$ (or its projection onto our brane, $\vec x_0$) as a function
of time. Left panel: the red thick line represents $a(t))$ under the effect of a 4-dimensional Newtonian force $F_4 (a)$; On the other hand, the blue thin lines
represent the brane-to-brane motion under the effect of $F_{\rm BB} (a,.d)$ for  $d = \pi$ (solid blue); $d = \pi/2$ (dashed blue); $d = \pi/4$ (dotted blue). 
Notice that Newtonian motion ends with a collision of the two masses in $t \simeq 390$ s.
Right panel: the red solid line represents Newtonian motion for a mass $m^\prime$ constrained on an incline plane at a minimal distance $d = \pi$ from $m$. 
The blue dashed line is the brane-to-brane motion under the effect of $F_{\rm BB}(a, \pi)$
In both panel the initial conditions are: $a_0 = 5$ ({\em i.e} $r_0 = 50$ $\mu$m) and $\dot a_0 = 0$. In the left panel, $k = k_5$ trivially due to eq.~(\ref{eq:5dimNewtonconstant}).
In the right panel, we have fixed $F_4 (a, \pi) \equiv F_{\rm BB} (a, \pi)$.}
\label{fig:Liumovement}
\end{figure*}

In Fig.~\ref{fig:Liumovement} we show the time evolution of the position $a$ of a body of mass $m^\prime$ under the effect of the gravitational force induced by 
a body of mass $m$ located at the origin of our three-dimensional coordinate system for the two cases in which the two bodies obey the 4-dimensional Newton's law (in red) 
or the brane-to-brane force $F_{\rm BB} (a,d)$ (in blue). We start at a distance $a_0 = 5$, {\em i.e.} $r = 50$ $\mu$m, and an initial velocity $\dot{a}_0 = 0$ in both cases. 
The initial distance is large enough for the 4-dimensional Newtonian force being a good starting approximation (the couplings $k$ and $k_5$ are taken to be identical).
However, under the effect of the gravitational force,
we see in the left panel of Fig.~\ref{fig:Liumovement} that the time evolution changes significantly. The 4-dimensional motion (thick red line) approaches $a = 0$ and 
stops in $t \simeq 390$ s, when the two bodies collide. On the other hand, the brane-to-brane motion reaches $a = 0$ and proceeds until $a = - a_0$ only to turn back 
and behave periodically like a pendulum. The period of the brane-to-brane motion depends on the distance of the two branes. We show three cases: $d = \pi$ (solid blue), 
$d = \pi/2$ (dashed blue) and $d = \pi/4$ (dotted blue), for which the period is $T \sim 1800$ s, $\sim 1600$ s and $\sim 1500$ s, respectively. Notice that the 4-dimensional 
motion follows the dashed blue line (corresponding to $d = \pi/2$) until crashing. This is a consequence of the particular shape of the brane-to-brane force: in 
Fig.~\ref{fig:5dpotential}(right panel)
we can see that for $d = \pi/2$ the brane-to-brane force is equivalent to the 4-dimensional Newton force down to distances of $a \sim 2$. On the other hand, 
for $d$ smaller the brane-to-brane force approaches the 5-dimensional force, that in that range of $a$ is stronger than the 4-dimensional one (and, thus, the resulting motion
is faster). For $d > \pi/2$ we have a slower motion, instead. 
In the right panel we show a slightly different situation: we consider the 4-dimensional Newtonian motion of a mass $m^\prime$ located on an inclined plane at minimal 
distance $d = \pi$ for the source $m$ of the gravitational field (red, solid line), and compare it with the motion of $m^\prime$ under the effect of the brane-to-brane force induced
by a source $m$ on a brane at a distance $d = \pi$ from our brane (blue, dashed line). 
The 5-dimensional coupling $k_5$ has been tuned such that the strength of $F_{\rm BB} (a, d) \equiv F_4 (a,d)$. We can see that the two motions are both periodic
and that the brane-to-brane motion is faster than the 4-dimensional motion, with a difference in the period of ${\cal O} (100)$ s.

\subsection{Orbital motion}
\label{sec:orbit}

It is now time to study the far more interesting case of two-dimensional motion. In this case, again, we can have open trajectories or orbits depending on the initial 
conditions. We will focus on the latter case, in which the mass $m$ at the source, the initial position and the initial angular velocity of the mass $m^\prime$ are tuned
such that a bounded orbit of $m^\prime$ around $m$ (or, more precisely, its projection onto our brane $\vec x_0$) is observed. 

Let's revise first the Newtonian case, where the equation of motion can be written as: 
\begin{equation}
m^\prime \ddot{{\vec r}} = {\vec F}_4 (\vec r) = - \vec \nabla V (r)  = - \frac{G_4 m \, m^\prime}{r^3} \, {\vec r} \, ,
\end{equation}
where $V (r)$ is the potential energy due to the gravitational field. The total energy is: 
\begin{equation}
\label{eq:energeticbalance}
{\cal E} = T + V = m^\prime \left \{ \frac{|{\vec v}|^2}{2} - \frac{G_4 m }{r} \right \} \, ,
\end{equation}
where $T$ is the kinetic energy of $m^\prime$. Writing the velocity in radial coordinates, we have:
\begin{equation}
{\vec v} = \dot{r} \, {\vec e}_r + r \dot{\theta} \, {\vec e}_\theta \, ,
\end{equation}
where $({\vec e}_r, {\vec e}_\theta)$ are two unit, orthogonal, vectors that define the position of $m^\prime$ at time $t$ in polar coordinates. 
Expressed in cartesian coordinates, ${\vec e}_r = (\cos \theta, \sin \theta)$ and ${\vec e}_\theta = (- \sin \theta, \cos \theta)$. 
In this basis, the acceleration becomes:
\begin{equation}
\dot{\vec v} = \left ( \ddot{r} - r \dot{\theta}^2 \right ) \, {\vec e}_r 
                              + \left ( r \ddot{\theta} + 2 \dot{r} \, \dot{\theta}  \right )  \, {\vec e}_\theta \, .
\end{equation}

It is now trivial to write a system of equations of motion for the mass $m^\prime$ in polar coordinates: 
\begin{equation}
\label{eq:newtonianeom}
\left \{ 
\begin{array}{lll}
\ddot{r} - r \, \dot{\theta}^2 &=& - \frac{G_4 m}{r^2} \\
 \\
r \, \ddot{\theta} + 2 \dot{r} \, \dot{\theta} & = & 0
\end{array}
\right .
\qquad \longrightarrow \qquad
\left \{ 
\begin{array}{lll}
\ddot{a} - a \, \dot{\theta}^2 &=& - \frac{k}{a^2} \, , \\
\\
a \, \ddot{\theta} + 2 \dot{a} \, \dot{\theta} & = & 0 \, ,
\end{array}
\right .
\end{equation}
where we have introduced the adimensional length $a = r/R$ and $k$ has been defined as in the previous section. 

If we now replace the Newtonian 4-dimensional force with the brane-to-brane force we have: 
\begin{equation}
\label{eq:b2beom}
\left \{ 
\begin{array}{lll}
\ddot{a} - a \, \dot{\theta}^2 &=& - \frac{k_5 }{a^2 }  \, f_{\rm BB} (a,d)  \, , \\
\\
a \, \ddot{\theta} + 2 \dot{a} \, \dot{\theta} & = & 0 \, ,
\end{array}
\right .
\end{equation}
The second equation implies conservation of angular momentum both for a Newtonian or a brane-to-brane force, 
\begin{equation}
h(t) = r^2 (t) \, \dot {\theta (t)} = h_0 \, ,
\end{equation}
where $h_0$ is a constant of motion. Using this result, the radial equation can be written as:
\begin{equation}
\label{eq:eombranetobrane}
\ddot{a} - \frac{h_0^2}{a^3} = \left \{ \begin{array}{l} - \frac{k}{a^2} \, , \\ \\  - \frac{k_5}{a^2} \, f_{\rm BB} (a,d) \, , \end{array} \right . 
\end{equation}
for the Newtonian (above) and brane-to-brane (below) cases, respectively. We get different results in the two cases: for the Newtonian case, solutions of the first of eqs.~(\ref{eq:newtonianeom}) are conic sections. Possible trajectories are, then, hyperbolic, parabolic or elliptic. In all cases, they can be described by a simple function,
\begin{equation}
\label{eq:newtonianorbit}
r (\theta) = \frac{r_c}{1 - e \, \cos \theta} \, ,
\end{equation}
where $r_c = h_0^2/ G_4 m$ and the eccentricity $e$ is given by
\begin{equation}
e = \frac{r_a - r_p}{r_a + r_p} \, ,
\end{equation}
being $r_a$ and $r_p$ the largest ({\em apoapsis}) and smallest ({\em periapsis}) distances of $m^\prime$ from $m$, respectively. 
For $e= 0$, $r (\theta)$ describes a circular orbit, whereas for $e <1$ the orbit is elliptic. For $e \geq1$ the trajectory is open, being parabolic for $e = 1$ 
and hyperbolic for $e > 1$. The period of a closed orbit of $m^\prime$ around $m$ can be computed easily applying the third Kepler's law:
\begin{equation}
\label{eq:3rdKeplerLaw}
T_{\rm N} =  \pi \frac{(r_a + r_p)^{3/2}}{\sqrt{2 G_4 m}} \, .
\end{equation}

\begin{figure*}[ht]
\centering
\includegraphics[angle=0,width=1\textwidth]{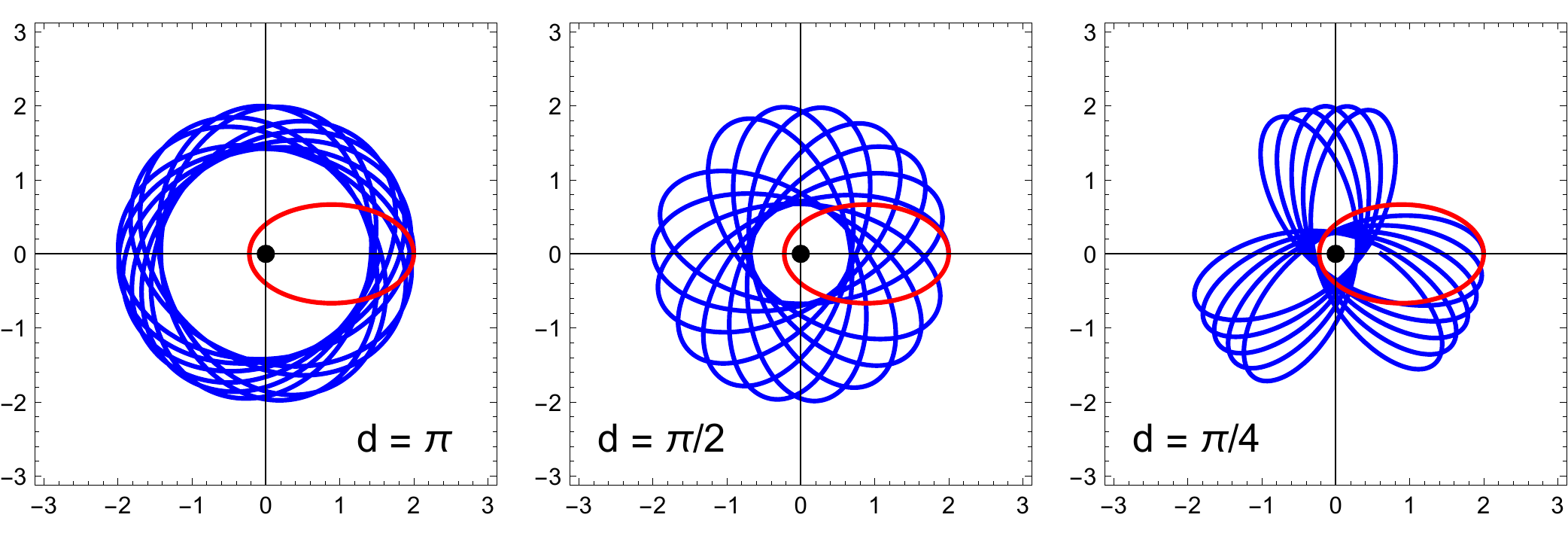}
\caption{\it  The trajectory of $m^\prime$ around the source $m$ located at (or whose projection in three dimensions lies at) $\vec x_0$, represented by a black dot. 
In red, we show the Newtonian elliptic orbit.
In blue, we show the motion under the effect of the brane-to-brane force when the two branes are at a distance $d = \pi$ (left panel), $\pi/2$ (middle panel) and $\pi/4$ (right panel), respectively. The initial conditions are as follows: $k_5 = k  = G_4 m/R^3 = 10^{-3}$ 1/s$^2$; $a_0 = 2$ ($r_0 = 20$ $\mu$m); $\dot{\theta}_0 = 5 \times 10^{-3}$ rad/s. }
\label{fig:orbits}
\end{figure*}

The results in the case of a brane-to-brane force are very different. Remember that, according to the Bertrand's theorem, closed orbits are only possible for central forces with a radial dependence of the form $1/r^2$ or $r$. Any deviation from these two possible functional dependences implies that the resulting orbits are not stable nor closed. 
A typical example of this is the general relativity correction to the orbit of Mercury: the leading post-Newtonian corrections are of the form $1/r^4$ and induce an observable precession of the perihelion of Mercury. This is precisely the case of the brane-to-brane force: the $r$-dependence of the (central) force field (either $F_5$ or $F_{\rm BB}$, depending if $d = 0$ or not) is not $1/r^2$.  As a consequence, we do not expect closed orbits (they may be bounded, though). This is indeed shown in Fig.~\ref{fig:orbits}, 
where we show the trajectory of $m^\prime$ around $\vec x_0$ (whose position is represented by a black dot at the origin) for $d= \pi$ (left panel), $\pi/2$ (middle panel) 
and $\pi/4$ (right panel), respectively. For the brane-to-brane motion, we have plotted (in blue) the first 100 revolutions of $m^\prime$ around $\vec x_0$, only.
In all cases, the initial conditions have been chosen such that the Newtonian orbit (depicted in red) is elliptic: $k = G_4 m/R^3 = 10^{-3}$ s$^{-2}$; $a_0 = 2$ ({\em i.e.} $r_0 = 20$
$\mu$m); $\dot{a}_0 = 0$; $\dot{\theta}_0 = 5 \times 10^{-3}$ rad/s ( {\em i.e.} $h_0 = 2$ $\mu$m$^2$ rad /s). The initial angle, $\theta_0$, can be chosen arbitrarily: we will 
fixed it at $\theta_0 = 0$. Since the initial radial velocity, $\dot{a}_0$, is set to be zero, the starting point $(a = a_0, \theta = 0)$ is necessarily either the periapsis or the apoapsis
of the orbit.

In all panels, we can see a significant precession of the periapsis that induces a rotation of the major axis of the orbit around $\vec x_0$. However, 
depending on the brane-to-brane distance $d$, the orbits can be very different even for the same choice of the initial conditions $a_0$, $\dot{a}_0$ and $\dot{\theta}_0$. 
In the left panel of Fig.~\ref{fig:orbits} (corresponding to $d = \pi$), for example, we can see that $m^\prime$ moves along nearly circular orbts with a slow 
counterclockwise precession of the periapsis. For $d = \pi/2$, orbits are elliptical, instead, whereas precession is still slow as for $d = \pi$. Eventually, for $d = \pi/4$, 
elliptical orbits are followed by fast nearly circular ones, and precession of the periapsis is fast, as the major axis rotate of approximately $45^\circ$ clockwise 
every two revolutions of $m^\prime$ around $\vec x_0$.

\subsection{Distinguishing a brane-to-brane from a Newtonian motion}
\label{sec:distinguish}

We want to study now the set of initial conditions for which is possible to distinguish a motion that is compatible with a Newtonian $1/r^2$ force from those that 
are clearly incompatible with that. To do this, we first compute the region of the parameter space for which we expect $m^\prime$ to orbit around a point. 
This is easily found computing the minimal angular velocity $\dot{\theta}_0$ for which a particle of mass $m^\prime$ at initial distance $r_0$ from $\vec x_0$ will
travel along an open trajectory. This is called the {\em escape velocity} and it can be computed looking when the kinetic energy exceeds the gravitational potential
in eq.~(\ref{eq:energeticbalance}), finding:
\begin{equation}
\dot{\theta}_0> \dot \theta_{0{\rm N}}^{\, {\rm esc}}  = \frac{\sqrt {2\, k }}{a_0^{3/2}}
\end{equation}
for a Newtonian potential, and 
\begin{equation}
\dot{\theta}_0> \dot \theta_{0{\rm BB}}^{\, {\rm esc}} = \frac{\sqrt {2\, k_5 }}{a_0^{3/2}} \, \left \{ \frac{\sinh a}{\cosh a - \cos d } \right \}^{1/2}
\end{equation}
for a brane-to-brane potential, respectively. In order to have an orbit (something that permits to study the geometrical properties of the trajectory over a long period of time) 
we must thus choose $\dot{\theta}_0$ and $a_0$ such that they would not violates the escape velocity bound. After checking this condition, 
we can measure the characteristics of the orbit.  Several features distinguish a Newtonian orbit from a non-Newtonian one. 
We will restrict ourselves in this section to study three of them: 
\begin{itemize}
\item The minimal distance\footnote{In the simulations and in the plots showing our results, we use as input variable the physical distance $r_0$, and not
        the adimensional distance $a_0 = r_0/R$.}  from the source of the gravitational field, $r_{\rm min}$ ({\em i.e.} the periapsis $r_p$ for a Newtonian orbit);
\item The maximal distance from the source of the gravitational field, $r_{\rm max}$ ({\em i.e.} the apoapsis $r_a$ for a Newtonian orbit);
\item The time it takes to $m^\prime$ to make a $2 \pi$-revolution around the source of the gravitational field, $T_{\rm BB}$ ({\em i.e.} the period $T_{\rm N}$ computed 
in eq.~(\ref{eq:3rdKeplerLaw}) for a Newtonian orbit).
\end{itemize}
Notice that $T_{\rm BB}$ is a quantity that should be easy to measure experimentally putting an electronic trigger at $\theta = 0$ ({\em e.g.} a laser beam can be sent along the $\theta = 0$ direction either to or from the source of the gravitational field, and when $m^\prime$ crosses the beam, thus interrupting it, a signal can be sent to a clock to measure the time lapse). Other possible definitions of $T$ for a non-closed orbit (such as the time it takes to $m^\prime$, starting at the maximal distance from $m$, to reach again the maximal distance, for example),  are not as easy to measure experimentally and will be therefore discarded.
Other geometrical features of the orbit could be used to distinguish the two models: for example, as it will be shown later, the precession of the periapsis is a characteristic feature
of non-Newtonian motion. However, without a specific description of the experimental setup used to measure this feature it is not easy to define an observable that can quantify
the amount of precession. For this reason, we have restricted ourselves in this Section to the limited but sufficient measurement of minimal and maximal distance of $m^\prime$ from 
$\vec x_0$. 

Consider now the following {\em gedanken} experiment: a particle of mass $m^\prime$ onto our brane ({\em here}) is put into motion around a gravitational source of mass $m$ 
that is located onto a parallel brane ({\em there}) at a distance $d$ from our brane. Clearly, we cannot "see" the source of the gravitational potential, as it may emit and absorb photons only in the other brane and it can be felt on our brane only gravitationally (for this reason the experiment is only a {\em gedanken} experiment). 
Still, we can put the particle of 
mass $m^\prime$ into motion with a certain set of initial conditions and measure the characteristics of its orbit. Assume that we know the mass $m$ of the source, the distance 
$d$ of the two branes and the location of the projection of $m$ onto our brane, $\vec x_0$. We can then define a set of possible initial conditions 
$\left \{ r_0, \dot r_0, \dot{\theta}_0 \right \}$. For simplicity, we have chosen $\dot r_0 = 0$ throughout our simulation (this is always possible once the position 
of the projection of $m$ onto our brane is known, as we are assuming,
and it corresponds to a particular choice of a coordinate system such that $\dot r|_{r_0} = 0, \dot \theta|_{r_0} = \dot \theta_0$). In our simulation, 
$k_5 R^3 = G_5 m/ 4 R = 1$ $\mu$m$^3$/s$^2$, corresponding to $m \sim 10^{-7}$ g and $R = 10$ $\mu$m. With this input, the transverse size of orbits is typically in the
tens of microns range. We have considered three possible distances of the two branes: $d = \pi/4, \pi/2$ and $\pi$. At this point, we can generate a mock data set including three observables: $\left \{ r_{\rm min}, r_{\rm max}, T_{\rm BB} \right \}$. The question to ask is:
is it possible to reproduce the data with a Newtonian potential? We have performed, therefore, a fit to the mock data using a Newtonian potential with only three free parameters,
$\left \{k, r_{0{\rm N}}, \dot \theta_{0{\rm N}} \right \}$, from which the Newtonian observable list $\left \{ r_p, r_a, T_{\rm N} \right \}$ can be univocally derived
using eqs.~(\ref{eq:newtonianorbit}) and (\ref{eq:3rdKeplerLaw}). 

As a first step, we have tried to fit the data using only  two geometrical information of the orbit, {\em i.e.} the minimum and maximum distance of $m^\prime$ from the source of the gravitational field, $r_{\rm min}$ and $r_{\rm max}$. In the case of a Newtonian potential, these two quantities correspond, as we have reminded above, to the periapsis $r_p$ 
and the apoapsis $r_a$, respectively. Having only two data points to fit, we have used a two-variables $\chi^2$: 
\begin{eqnarray}
\label{eq:chi22dofs}
\chi^2_{\rm 2 obs} &=& \min_{k, r_{0{\rm N}}, \dot \theta_{0{\rm N}}}  
\left \{ \frac{\left [ r_p (k, r_{0{\rm N}}, \dot \theta_{0{\rm N}} ) - r_{\rm min}(k_5, d; r_0, \dot \theta_0) \right ]^2}{\sigma_r^2} \right . \nonumber \\
&+& \left . \frac{\left [ r_a (k, r_{0{\rm N}}, \dot \theta_{0{\rm N}} ) - r_{\rm max}(k_5, d;  r_0, \dot \theta_0) \right ]^2}{\sigma_r^2}  \right \} \, .
\end{eqnarray}
In the computation of $\chi^2_{\rm 2obs}$ we have assumed that
the measurements of the minimum and maximum distance of $m^\prime$ from the source of the gravitational field
 are gaussian distributed variables with variance $\sigma_r = 1$ $\mu$m. Remember that, for our choice of $k_5$, orbits have a typical size of tens of microns. Therefore, 
the relative error on the measurement of a distance ranges from 10\% (for small orbits) to 1\% (for large orbits). It is probably possible to measure distances at this
length scale with an error better than 1 $\mu$m. However, we consider it a conservative choice. As a second step, we have added the dynamical information regarding 
the measurement of the period $T$ (defined above). For a Newtonian orbit, this is not an independent variable, as it can be univocally determined using the third Kepler's law
knowing $r_p, r_a$ and $k$. For this reason, adding this piece of information to the fit can be a powerful tool to distinguish between a truly Newtonian orbit and a manifestly
non-Newtonian one. In this case, we fit our mock data using a $\chi^2$ with three observables:
\begin{eqnarray}
\label{eq:chi23dofs}
\chi^2_{\rm 3 obs} &=& \min_{k, r_{0{\rm N}}, \dot \theta_{0{\rm N}}}
\left \{ \frac{\left [ r_p (k, r_{0{\rm N}}, \dot \theta_{0{\rm N}} ) - r_{\rm min}(k_5, d; r_0, \dot \theta_0) \right ]^2}{\sigma_r^2} 
\right . \nonumber \\
&+&  \frac{\left [ r_a (k, r_{0{\rm N}}, \dot \theta_{0{\rm N}} ) - r_{\rm max}(k_5, d;  r_0, \dot \theta_0) \right ]^2}{\sigma_r^2} \\
&+&  \left . \frac{\left [ T_{\rm N } (k, r_{0{\rm N}}, \dot \theta_{0{\rm N}} ) - T_{\rm BB}(k_5, d;  r_0, \dot \theta_0) \right ]^2}{\sigma_T^2} 
\right \} \, . \nonumber
\end{eqnarray}
Also in this case, we assume that the measure of the time required for $m^\prime$ to complete a $2 \pi$-revolution around the source of the gravitational field is a gaussian
distributed variable with variance $\sigma_T = 1$ s. Typical periods $T_{\rm BB}$ in our mock data range from hundreds to thousands of seconds. Therefore, this error on the measurement of a period corresponds to a 0.1\%-1\% error, approximately. Notice that this is a {\em very} conservative choice, given the state-of-art capability to measure time lapses. However, in most cases it will be enough. 

\begin{figure*}[ht]
\centering
\includegraphics[width=14cm]{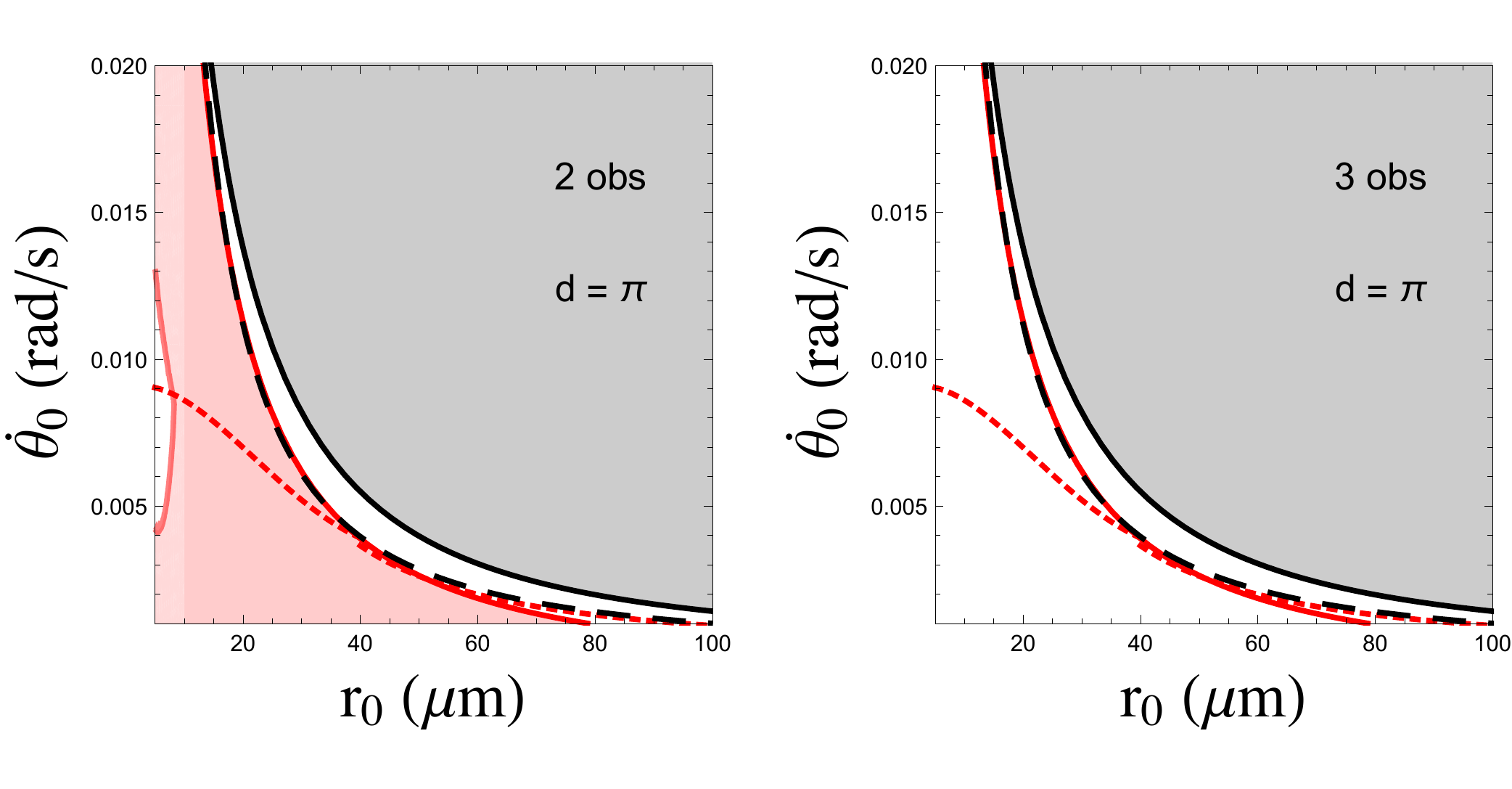} 
\caption{\it  
The $p$-value contour delimiting the region of the ($r_0, \dot \theta_0$) parameter space for which a Newtonian force can or cannot reproduce
data corresponding to the orbit of an object of mass $m^\prime$ induced by the gravitational field generated by a mass $m$ located on a brane at distance $d = \pi$.
The red-shaded (white) area corresponds to the region for which the $p$-value is above (below) 0.05, {\em i.e.} in this region 
the null hypothesis cannot (can) be rejected. The grey-shaded area corresponds to the region for which $m^\prime$ "escapes" to the force $F_{\rm BB}$ generated by $m$.
 Left panel: fit performed using measurement of two observables ($r_{\rm min}$ and $r_{\rm max}$), eq.~(\ref{eq:chi22dofs}). 
 Right panel: fit performed using measurement of three observables ($r_{\rm min}$, $r_{\rm max}$ and $T$), eq.~(\ref{eq:chi23dofs}). 
 In both panels, the black dashed line (red dotted line) represents the value of $\dot \theta_0$ for which the orbit is circular, 
$\dot \theta_{0, {\rm N}}^{\rm \, crit}$ ($\dot \theta_{0, {\rm BB}}^{\rm \, crit}$).
 }
\label{fig:fittopi}
\end{figure*}

\begin{figure*}[ht]
\centering
\includegraphics[width=14cm]{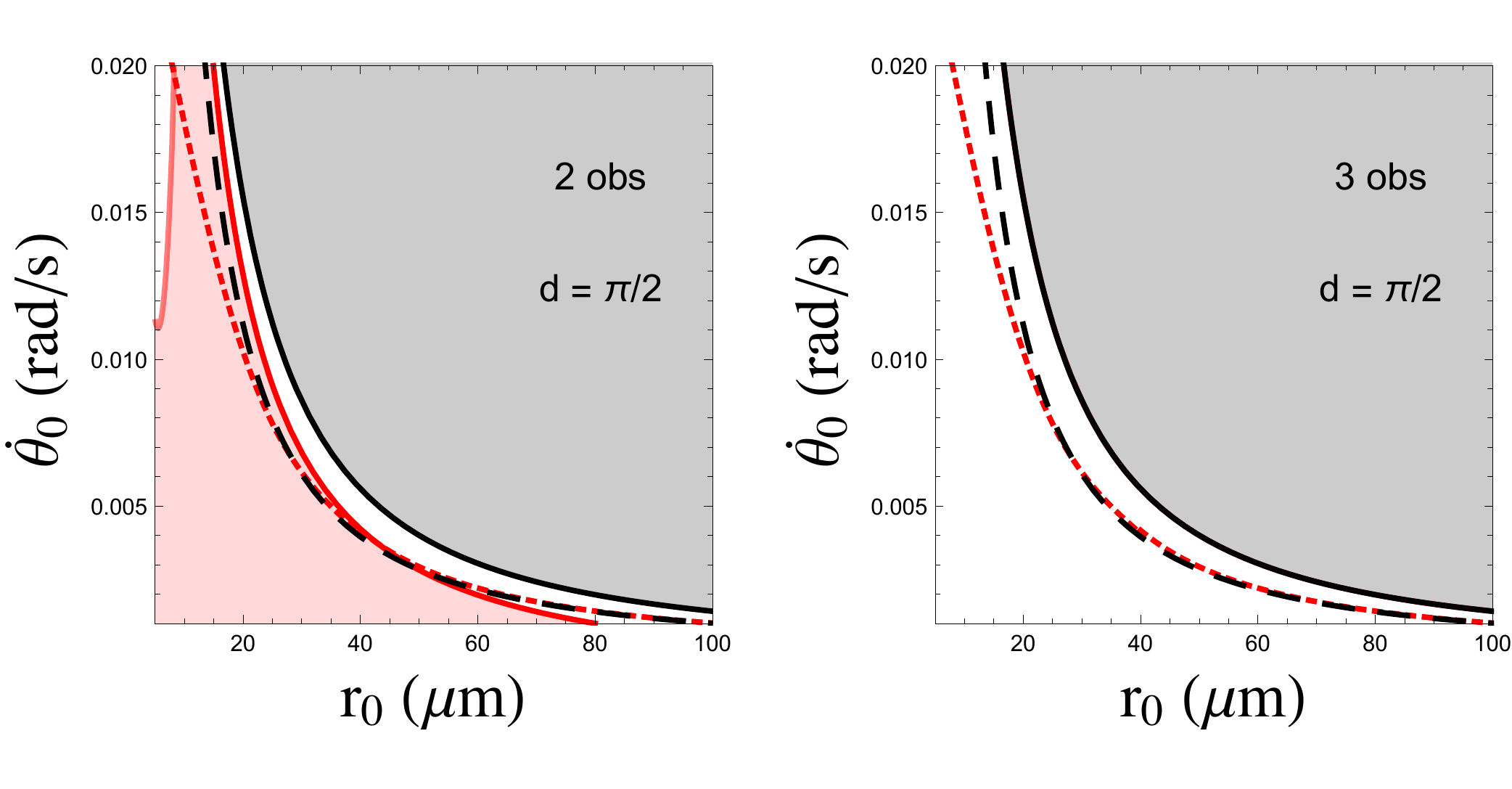} 
\caption{\it  The same as in Fig.~\ref{fig:fittopi} for $d = \pi/2$.}
\label{fig:fittopi2}
\end{figure*}

\begin{figure*}[ht]
\centering
\includegraphics[width=14cm]{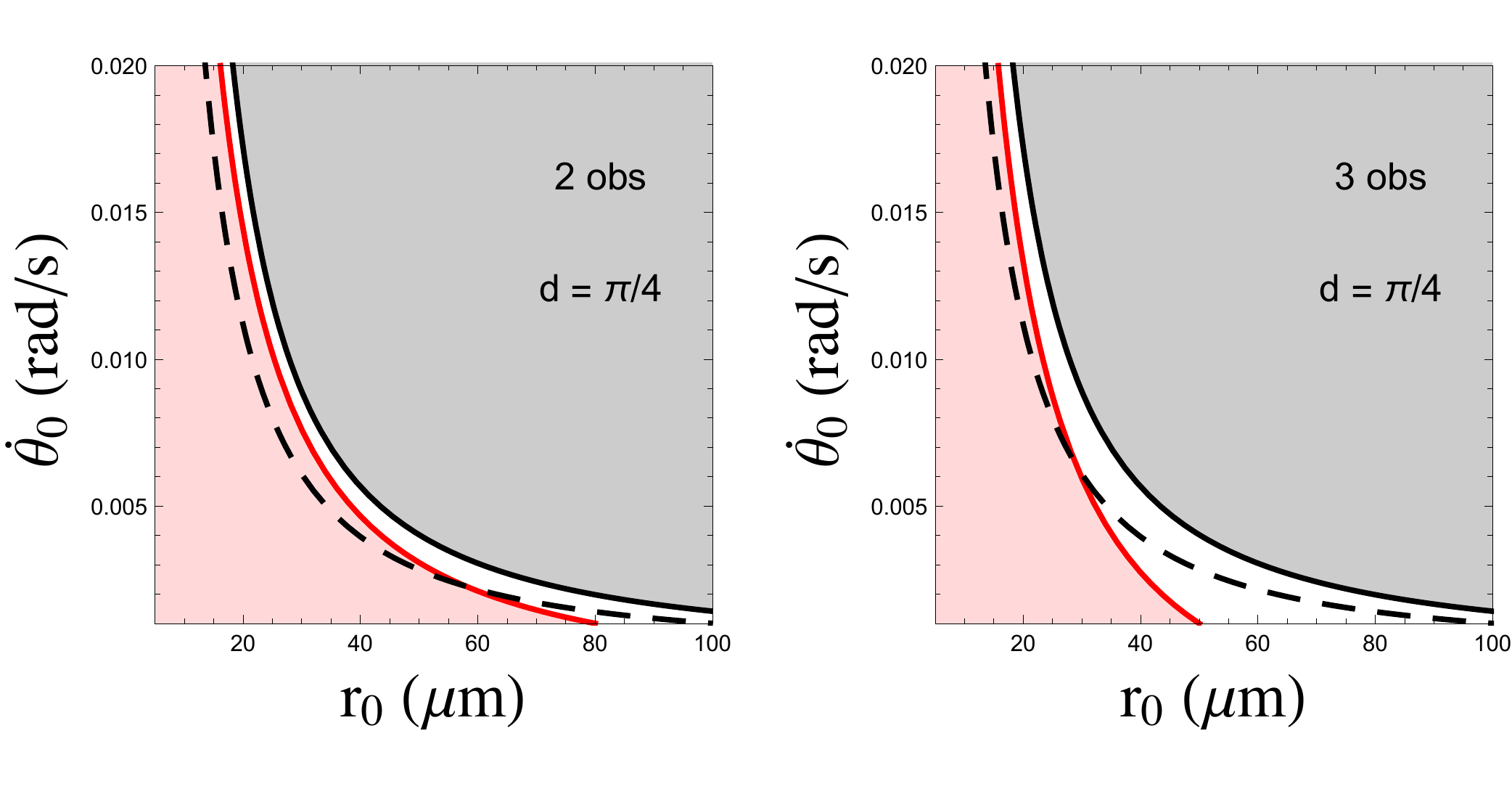} 
\caption{\it  The same as in Fig.~\ref{fig:fittopi} for $d = \pi/4$.}
\label{fig:fittopi4}
\end{figure*}

What we are doing here, {\em i.e.} fit "experimental" data with a theoretical model asking if the model is able to reproduce the data, is a hypothesis test.
The hypothesis $\bf{H}$ that we test is that {\em data are distributed so as to reproduce some geometrical and dynamical features of a Newtonian orbit}
(in statistics, this is called the {\em null hypothesis}). In order to accept or reject this hypothesis, we adopt the following strategy \cite{Agashe:2014kda}: 
\begin{enumerate}
\item We first minimize the $\chi^2$ functions defined in either eq.~(\ref{eq:chi22dofs}) or (\ref{eq:chi23dofs}), obtaining $\chi^2_{\rm min}$.
         If the measured observables behave as gaussian variables, then $\chi^2_{\rm min}$ is distributed according the $\chi^2$ probability density function, 
         $f \left ( \chi^2, n_d \right ) $, with $n_d$ the number of degrees of freedom\footnote{
         Usually, the number of degrees of freedom of a $\chi^2$ fit is $n_d = N - P$, where $N$ is the number of data points and $P$ the number of fitting
         variables. However, this is strictly true ONLY when the model that we use to fit the data is linear, 
         {\em i.e.} $X( \vec n, \vec \theta) = \theta_1 A_1 (\vec n) + \dots \theta_P A_P (\vec n) $, where $\vec n$ ($n = 1, \dots, N$) is the data vector, $\vec \theta$ is the
         free parameters vector  ($\vec \theta = 1, \dots, P$) and $\vec A (\vec n)$ is a basis of functions that depend on the data set. If the functions that form the
         basis are independent between themselves, then $n_d = N - P$ (otherwise, in general one would get $N-1 > n_d > N - P$). However, when the model 
         that we use to fit the data is non-linear , $n_d$ cannot be computed straightforwardly (see Ref.~\cite{Andrae:2010gh}
         and refs. therein for some example on this subject). This is, indeed, 
         our case, as eqs.~(\ref{eq:newtonianorbit}) and (\ref{eq:3rdKeplerLaw}) imply non-linear
         relations between the fit parameters $r_{0{\rm N}}, \dot \theta_{0{}\rm N}$ and $k$.
         For this reason, since we want to draw qualitative conclusions on the capability of a Newtonian model
         to fit data produced by a brane-to-brane force, we will fix $n_d = 1$ in our simulations. 
         }.
         The $\chi^2$ p.d.f. gives the probability to get a certain value of $\chi^2_{\rm min}$ when performing a $\chi^2$ fit to a set of data, given that the data are
         gaussian distributed and that the model used to fit the data is correct.
\item We can then compute the $p$-value:
\begin{equation}
\label{eq:pvalue}
p = \int_{\chi^2_{\rm min}}^\infty \, d \chi^2 \, f \left ( \chi^2, 1 \right )  \, .
\end{equation}
         The $p$-value, as defined above, computes the area of the tail of the $\chi^2$ p.d.f. If $p$ is small, then $\chi^2_{\rm min}$ is large and the goodness-of-fit is 
         poor ({\em i.e.} it would be unlikely that rejecting the hypothesis $\bf{H}$ be a wrong choice). A typical value below which the discrepancy between the hypothesis $\bf{H}$ 
         and the data is considered to be significant is $p = 0.05$. 
         \item
         We eventually draw contours for $p = 0.05$ in the $(r_0, \dot \theta_0)$-plane. The results of our hypothesis test are shown in Figs.~\ref{fig:fittopi}, \ref{fig:fittopi2} and   
         \ref{fig:fittopi4} for $d=\pi$, $d = \pi/2$ and $d = \pi/4$, respectively. 
\end{enumerate}

In all figures, the region of the parameter space for which the fit to data using a Newtonian potential is considered to be good ({\em i.e.} where $p > 0.05$) is represented by the light red-shaded area. The region  of the parameter space for which we have an open trajectory ({\em i.e.} where $\dot \theta_0 > \dot \theta_{0, {\rm BB}}^{\rm \, esc}$) is gray-shaded. Eventually, black dashed and red dotted lines represent the choice of initial conditions for which a Newtonian (non-Newtonian) orbit is circular ({\em i.e.} $r_{\rm min} = r_{\rm max}$). Let's call these lines as $\dot \theta_{0 {\rm N}}^{\rm \, crit}$ and $\dot \theta_{0 {\rm BB}}^{\rm \, crit}$, respectively.

Consider first the case of $d = \pi$, shown in Fig.~\ref{fig:fittopi}. Using only information from the measurement of $r_{\rm min}$ and $r_{\rm max}$ (left panel), the result of a fit to data under the hypothesis that data should reproduce a Newtonian orbit is very good in, approximately, all of the allowed parameter space ({\em i.e.} in the region for which we expect a non-open trajectory). There are two regions for which the fit is not good, and therefore rejecting the hypothesis is unlikely 
to be wrong. The first one is a narrow strip near the bound where trajectories become open. Notice that the grey shaded area represents the region
of the parameter space for which $m^\prime$ escapes to the gravitational force $F_{\rm BB}$ generated by the source $m$ located on a distant
brane. For values of the parameters near the escape line, the time needed to make a $2 \pi$-revolution become longer and the orbit is very long
(as it happens for trans-plutonian objects in the Solar System). On the other hand, the escape line for a Newtonian force (not plotted) lies within the grey shaded area, 
and orbits in the Newtonian case are shorter and faster. For this reason, the fit in this region gives generically a small $p$-value. The second
region where the fit is not good corresponds to low $r_0$ and $\dot \theta_0 \sim \dot \theta_{0{\rm BB}}^{\rm \, crit}$. 
This happens since for this particular choice of the input values $(r_0, \dot \theta_0)$ the data describes a nearly circular orbit (see the left panel of Fig.~\ref{fig:orbits}), 
whereas a Newtonian potential would try to fit them with a hugely elliptical one 
(as it can be seen looking at the black dashed line, for which  $\dot \theta_{0{\rm N}}^{\rm \, crit} \gg \dot \theta_{0{\rm BB}}^{\rm \, crit}$ for $r_0 \sim R$ and this value of $d$). 
Below the red dotted line the BB-orbits are elliptical, too, and the Newtonian model is able to mimic the data.
The results are quite different when we introduce information from the measurement of the time required to make a $2 \pi$-revolution,  $T_{\rm BB}$ (right panel):
in this case,  a Newtonian fit to the data gives an extremely small $p$-value in all the parameter space. We conclude that for $d = \pi$,  the measurement of the period 
with an error $\sigma_T = 1$ s is necessary (and sufficient) to exclude that the observed trajectory is Newtonian. 

Consider now the case of $d = \pi/2$, Fig.~\ref{fig:fittopi2}. The fit to two observables (left panel) is very similar to that at $d = \pi$. The only
difference is that the critical line $\dot \theta_{0, {\rm BB}}^{\rm \, crit}$ (red dotted line) is very similar to the Newtonian critical
line $\dot \theta_{0, {\rm N}}^{\rm \, crit}$ (black dashed line) for most of the values of $\dot \theta_0$ in the figure; as a consequence, the region for which the fit 
is bad at low $r_0$ moves upward (where the difference between the two lines increases). As for $d = \pi$, in the right panel we can see
that, after including the measurement of the time needed to make a $2 \pi$-revolution, the Newtonian fit is able to reproduce the data in all of the considered region 
of the initial conditions parameter space.

Consider, eventually, the case of $d = \pi/4$, Fig.~\ref{fig:fittopi4}. The fit to two observables (left panel) shows that a Newtonian potential is able to fit the mock data
in all of  the considered parameter space. Notice that, in this case, the brane-to-brane and the Newtonian critical lines $\dot \theta_{0, {\rm BB}}^{\rm \, crit}$ and
$\dot \theta_{0, {\rm N}}^{\rm \, crit}$ coincide for $\dot \theta_0 \in [0.001,0.020]$ (they start to differ for larger values of $\dot \theta_0$). 
For this reason, no area at low $r_0$ with a poor fit can be found. Once the measurement
of the $2 \pi$-revolution time lapse is taken into account, we are still not able to distinguish the two models in most of the parameter space. 
It is interesting to stress, however, that a region for which a Newtonian fit cannot explain the observed data is found at large $r_0$, low $\dot \theta_0$. This is in apparent
contradiction with eqs.~(\ref{eq:Yukawapotential}) and (\ref{eq:Yukawapotential2}), from which we can see that, for large $r_0$, $V_5$ should approach a Newtonian potential
exponentially. This is because, once an angular momentum is included, in the considered range of $r_0$ the dynamics
induced by a Newtonian force still differs from that induced by $F_{\rm BB}$ (and, thus, $T_{\rm BB} \neq T_{\rm N}$). Since $\sigma_T = 1$ s, the difference
in the revolution times is large enough to invalid the null hypothesis. 
On the other hand, for larger values of $r_0$ we expect that the distinction between the two models be no longer possible.

As a last comment, we have checked that for $d = \pi/4$ it is possible to reject the Newtonian hypothesis in the whole considered
parameter space if the error on the measurement of the time needed to perform a $2 \pi$-revolution of $m^\prime$ around $\vec x_0$ is lowered. 
This can be done using $\sigma_T = 0.1$ s, certainly nothing exceedingly difficult to achieve given the state-of-art electronics. 

\section{Two bodies on the same brane}
\label{sec:d0}

We have seen in the previous section that, once the mass $m^\prime$ acquires a small angular velocity, the time needed to perform a $2 \pi$-revolution around the 
projection of the source of the gravitational field $m$ can differ significantly between a Newtonian and a brane-to-brane motion. This is still true even when the two masses lie onto
the same brane, {\em i.e.} in the case $d = 0$. For this reason, in this section we will study in more detail this case, that can be of direct relevance to improve the 
bounds on deviations from the $1/r^2$ Newton's law.

The problem we want to study is that of a classical two-body gravitational system with a "planet" $P$ with mass $m \sim 10^{-7}$ g and a "satellite" $S$
with mass $m^\prime \sim 10^{-9}$ g, such that we can neglect the motion of $P$ under the effect of $S$. 
 As we have seen in the previous section, with this choice of masses, the typical orbit of $S$ around $P$ has a radius of 
tens to hundreds of microns (depending on the initial position $r_0$ and on the initial angular velocity $\dot \theta_0$). We consider, therefore, a "laboratory"
with a size of 1 mm$^2$. The source should be made of a compact material, in order to reduce its size: for a spherical iron source of mass $m = 10^{-7}$ g, the radius
is $r_P = 14.5 $ $\mu$m; for a platinum source with the same mass, $r_P = 10.3$ $\mu$m. On the other hand, a satellite $S$ of mass $m^\prime = 10^{-9}$ g
has a typical size ranging from 2 to 3 $\mu$m, depending on the material\footnote{In principle, to reduce backgrounds due to electrical forces between $P$ and $S$, the
satellite should be an insulator. However, alternative choices could be made, depending on the setup adopted (see {\em e.g.} Refs.~\cite{Geraci:2010,Yin:2013lqa,Geraci:2014gya}).}. 
To get an idea, the ratios of masses and radii of $S$ to $P$ are very similar to the corresponding ratios for the Moon and the Earth. 
The relative distance between $S$ and $P$ that we are considering, on the other hand, is much shorter than the distance between the Earth and the Moon. 
The satellite $S$ remains in orbit around the planet $P$ because the range of angular velocity that we are dealing with is much larger than the angular velocity
of the Moon around the Earth. 
The first difference between the $d= 0$ and $d \neq 0$ cases is that the potential diverges when $m^\prime$ approaches the source
of the gravitational field. Taking into account the physical size of the source and of the satellite, we must choose the range of the initial conditions so as to avoid a 
collision between $P$ and $S$. We consider, therefore, the initial distance between the two bodies larger than in the case $d \neq 0$: $r_0 \in [100, 200]$ $\mu$m.
The range of angular velocities such that $S$ does not collide with $P$ and does not escape from it is rather narrow for this choice of $r_0$: 
$\dot \theta_0 \in [1.5 \times 10^{-4}, 1.5 \times 10^{-3}]$ rad/s (notice that the Moon angular velocity around the Earth is $2.66 \times 10^{-6}$ rad/s). For a typical choice
of initial conditions within the range give above, $r_0 = 190$ $\mu$m and $\dot \theta_0 = 1.8 \times 10^{-4}$ rad/s, we get a very eccentric Newtonian orbit, $e = 0.775$, to be compared with the nearly circular Moon-Earth orbit, for which $e = 0.0549$.

\begin{figure*}[ht]
\centering
\includegraphics[width=14cm]{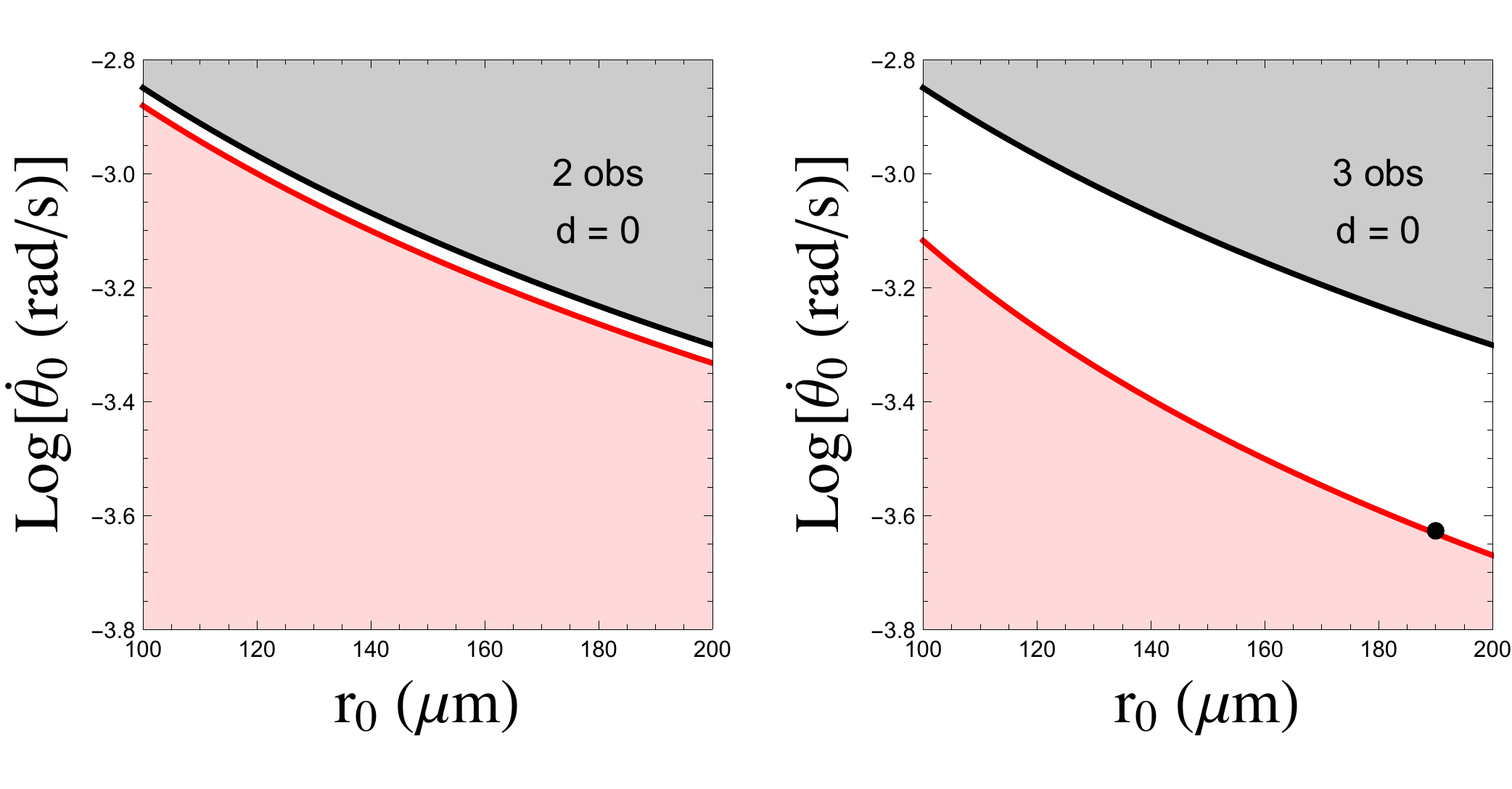} 
\caption{\it  The $p$-value contour delimiting the region of the ($r_0, \dot \theta_0$) parameter space for which a Newtonian force can or cannot reproduce
data corresponding to the orbit of an object of mass $m^\prime$ induced by the gravitational field generated by a mass $m$ located on the same brane
({\em i.e.} $d = 0$).
The red-shaded (white) area corresponds to the region for which the $p$-value is above (below) 0.05, {\em i.e.} in this region 
the hypothesis cannot (can) be rejected at 95 \% CL. The grey-shaded area corresponds to the region for which $m^\prime$ "escapes" to the force $F_5$ generated by $m$.
 Left panel: fit performed using measurement of two observables ($r_{\rm min}$ and $r_{\rm max}$), eq.~(\ref{eq:chi22dofs}). 
 Right panel: fit performed using measurement of three observables ($r_{\rm min}$, $r_{\rm max}$ and $T_5$), eq.~(\ref{eq:chi23dofs}). The black dot shown in this panel
  represents the initial conditions choice $r_0 = 190$ $\mu$m, $\dot \theta_0 = 1.8 \times 10^{-4}$ rad/s discussed below.
 }
\label{fig:fitto0}
\end{figure*}

As in the previous section, we have performed a statistical analysis of the goodness of a Newtonian fit
to mock data produced using the 5-dimensional force $F_5$. Our results are shown in Fig.~\ref{fig:fitto0}. Again, the grey-shaded area represents the region 
for which $S$ escapes the gravitational field of $P$, whereas the light red-shaded area represents the region of the parameter space for which rejecting the Newtonian
hypothesis is likely to be wrong ({\em i.e.} the region for which $p > 0.05$). The left panel represents a fit to only two observables, $r_{\rm min}$ and $r_{\rm max}$, 
whereas the right panel includes the information on the time needed for $S$ to perform a $2 \pi$-revolution around $P$, $T_5$. In order to present the narrow region 
of allowed angular velocities, we have shown the vertical axis in logarithmic scale. Notice that, for simplicity, we have considered in our numerical simulations 
only the case in which the compactification radius is $R = 10$ $\mu$m. 

\begin{figure*}[ht]
\centering
\includegraphics[width=14cm]{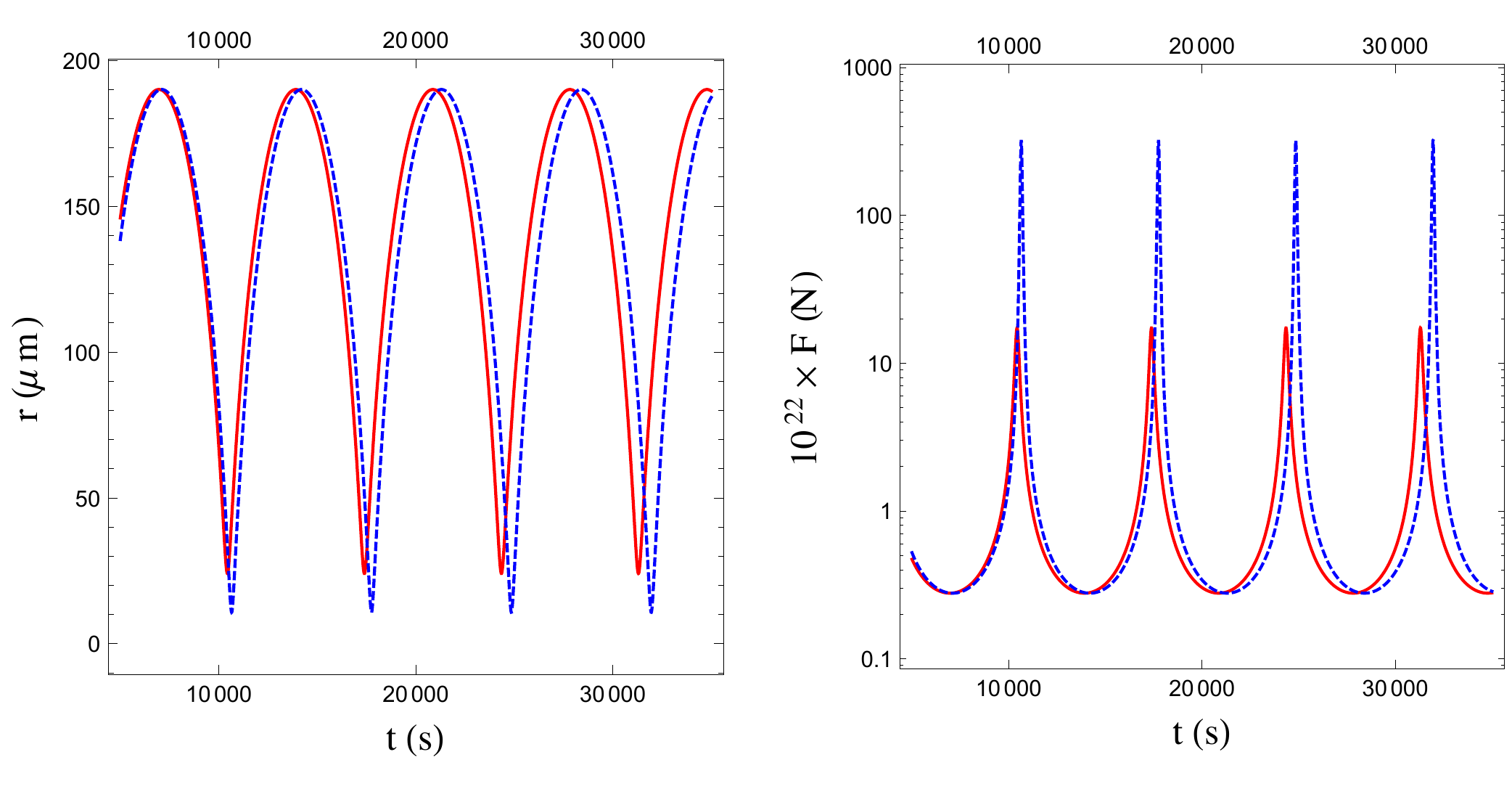} 
\caption{\it Distance of S from P and gravitational force at the S position over time for the initial conditions choice $r_0 = 190$ $\mu$m and 
$\dot \theta_0 = 1.8 \times 10^{-4}$ rad/s.
Left panel: the distance of the satellite S from the planet P as a function of time.
Right panel: the gravitational force felt by the satellite S under the effect of the planet P as a function of time, multiplied by a factor $10^{22}$. 
In both panels,  red, solid lines represent the Newtonian case, whereas blue, dashed lines represent the 5-dimensional case. 
 }
\label{fig:checkpointd0orbit1}
\end{figure*}

As we can see from the right panel of Fig.~\ref{fig:fitto0}, the information coming from the measurement of the time needed to perform a $2 \pi$-revolution of S around P
is necessary in order to distinguish the Newtonian orbit from the 5-dimensional one. Once this information is included, a white strip in the $(r_0, \dot \theta_0)$-plane for which
the distinction is possible emerges. In order to understand better why the two cases give significantly different results, we choose a representative point within the white
region of the $(r_0, \dot \theta_0)$-plane and study the main characteristics of the corresponding orbits. 
Consider, then, the case of $r_0 = 190$ $\mu$m, $\dot r_0 = 0$ and $\dot \theta_0 = 1.8 \times 10^{-4}$ rad/s, represented by a black dot in Fig.~\ref{fig:fitto0} (right panel). 
The dependence of the distance of S from P as a function of time for the Newtonian and the 5-dimensional cases are shown in the left panel of 
Fig.~\ref{fig:checkpointd0orbit1} in red, solid  (blue, dashed) lines, respectively. Notice that the plot doesn't show $t = 0$, for which necessarily $r_0$ coincides with
the apoapsis $r_a$ due to the initial condition choice. As we can see, the information concerning the distance of S from P is not
much inspiring: the maximum distance is always identical for the two cases, whereas the minimum distance of S from P 
(the periapsis, $r_p$) is a bit shorter for the 5-dimensional case with respect to the Newtonian case. We also notice a rather small shift in the time needed to regain the apoapsis after one revolution. In the right panel of the same figure we present, on the other hand, the gravitational force felt by S under the effect of P along its orbit (multiplied by
a convenient factor $10^{22}$). We can see that, when S reach its periapsis, the force in the 5-dimensional case can indeed be much larger than for the Newtonian case. For the particular choice of $r_0$ and $\dot \theta_0$ given above, we have that $F_N (r = r_{p,N}) = 17.3 \times 10^{-22}$ N whereas $F_5 (r_{p,5}) = 322.8 \times 10^{-22}$ N, {\em i.e.} 
approximately twenty times larger! 

\begin{figure*}[ht]
\centering
\includegraphics[width=14cm]{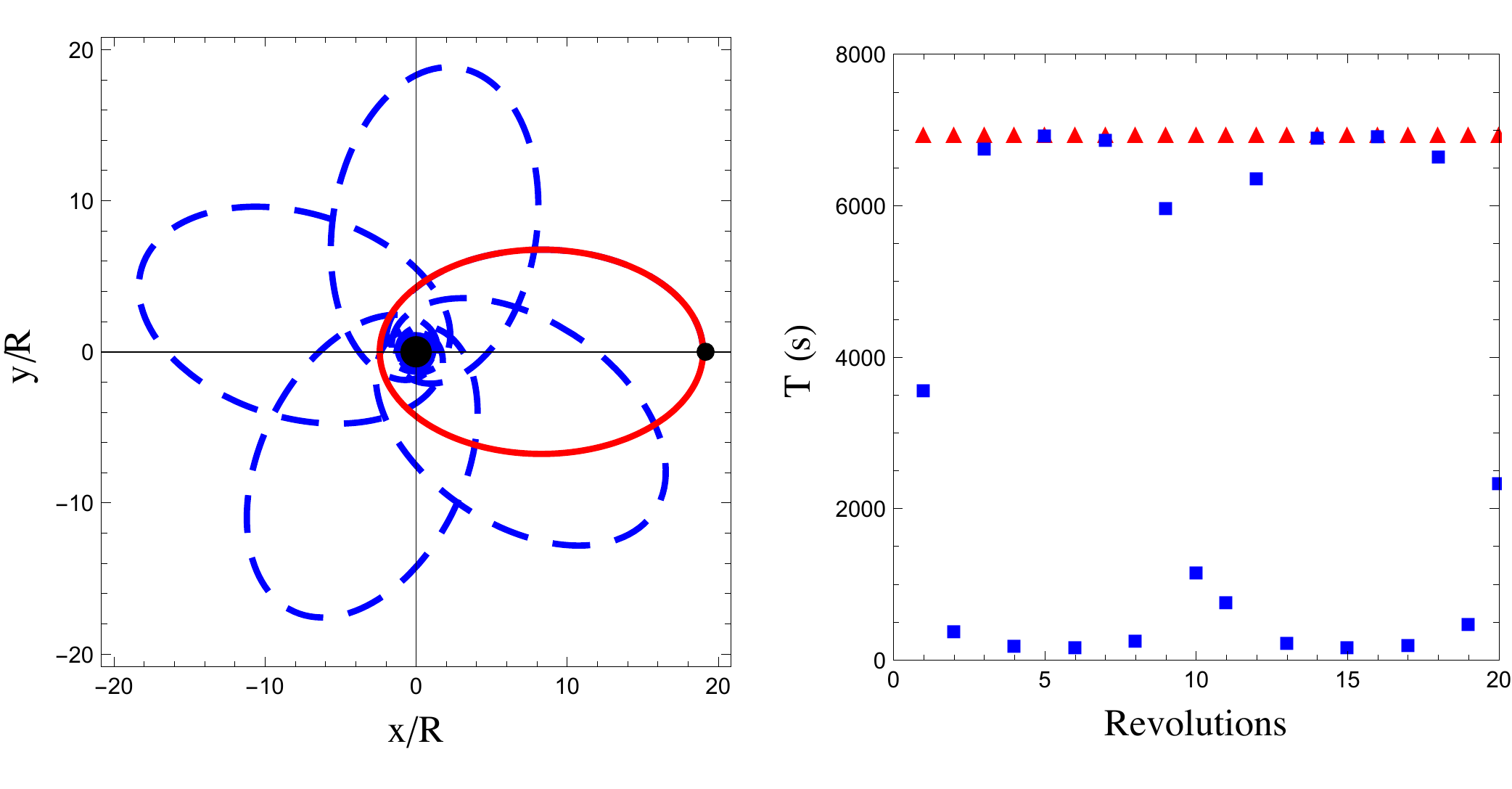} 
\caption{\it 
Left panel: the orbit of the satellite S (depicted by a black dot at $x_0/R = 19, y_0/R = 0$) around the planet P (at the center of the plot) in the orbit plane
for the initial conditions choice $r_0 = 190$ $\mu$m and $\dot \theta_0 = 1.8 \times 10^{-4}$ rad/s. The red , solid line
represents the Newtonian orbit, whereas the blue, dashed line represents the 5-dimensional orbit.
Right panel: the measurement of the time needed to perform $n$ $2 \pi$-revolutions of S around P for the same choice of initial conditions.
Red triangles represent the Newtonian case, whereas blue squares represent the 5-dimensional case. 
 }
\label{fig:orbitd0}
\end{figure*}

The impressive enhancement of the gravitational force at the periapsis alters completely the orbit of S around P. This is shown in Fig.~\ref{fig:orbitd0}, where the 
Newtonian orbit is represented as a red, solid line and the first ten (!) revolutions of S around P are shown by blue, dashed line. The black disk at the center of the plot represents the platinum source with a physical size $r_P /R = 1.03$, whereas the satellite is represented by a small black dot starting at a $r_0 = 190$ $\mu$m distance on the positive horizontal axis.  Notice that the angular velocity has been fine-tuned so that the 5-dimensional orbit never touches the source, {\em i.e.} the satellite S never
crashes onto the planet P. However, every time that S approaches its periapsis, the source P induces a gravitational slingshot on it, modifying completely its trajectory. The 5-dimensional orbit can be described as follows: after a first half-revolution that follows approximately the Newtonian trajectory, the gravitational force of P makes S perform a very fast and short circular orbit around P, only to regain an almost elliptical path
that eventually brings it to a new apoapsis, albeit with an approximate $90^\circ$ shift of the ellipse major axis with respect to the Newtonian orbit. This pattern: (1) a {\em long and slow, almost Newtonian, revolution},  followed by (2) a {\em short and fast, almost circular, one}, repeats until finally regaining (approximately) the initial position after ten revolutions, as shown in the Figure. It is clear that the 5-dimensional orbit is geometrically completely different from the Newtonian one. 
As we will see, the time needed to perform a revolution differs as well. 

In the right panel of Fig.~\ref{fig:orbitd0} we plot the times that S needs to perform a revolution around P.  In the Newtonian case, depicted by red triangles, every
revolution takes the same time, $T_N$, that for the particular choice of initial conditions given above is $T_N = 6956.8$ s, {\em i.e.} almost two hours! 
The blue squares represent, on the other hand, the revolution times in the 5-dimensional case, $T_{5n}$, where $n$ stands for the $n$-th $2 \pi$-revolution of S around P.
In this case, we can appreciate immediately the effect of the gravitational slingshot induced by the huge enhancement of the gravitational force at the periapsis 
in the 5-dimensional case with respect to the Newtonian case: revolution times approximately similar to those computed in the Newtonian case are followed
by much shorter revolution times, ranging from $\sim 100$ s to $\sim 1000$ s. It is this information that can be best used to distinguish the two cases and
to improve our present limits on the deviations from the $1/r^2$ Newton's law.

\section{Deviations from the $1/r^2$ Newton's law in 4-dimensions}
\label{sec:experiment}

\begin{figure}[ht]
\centering
\includegraphics[width=8cm]{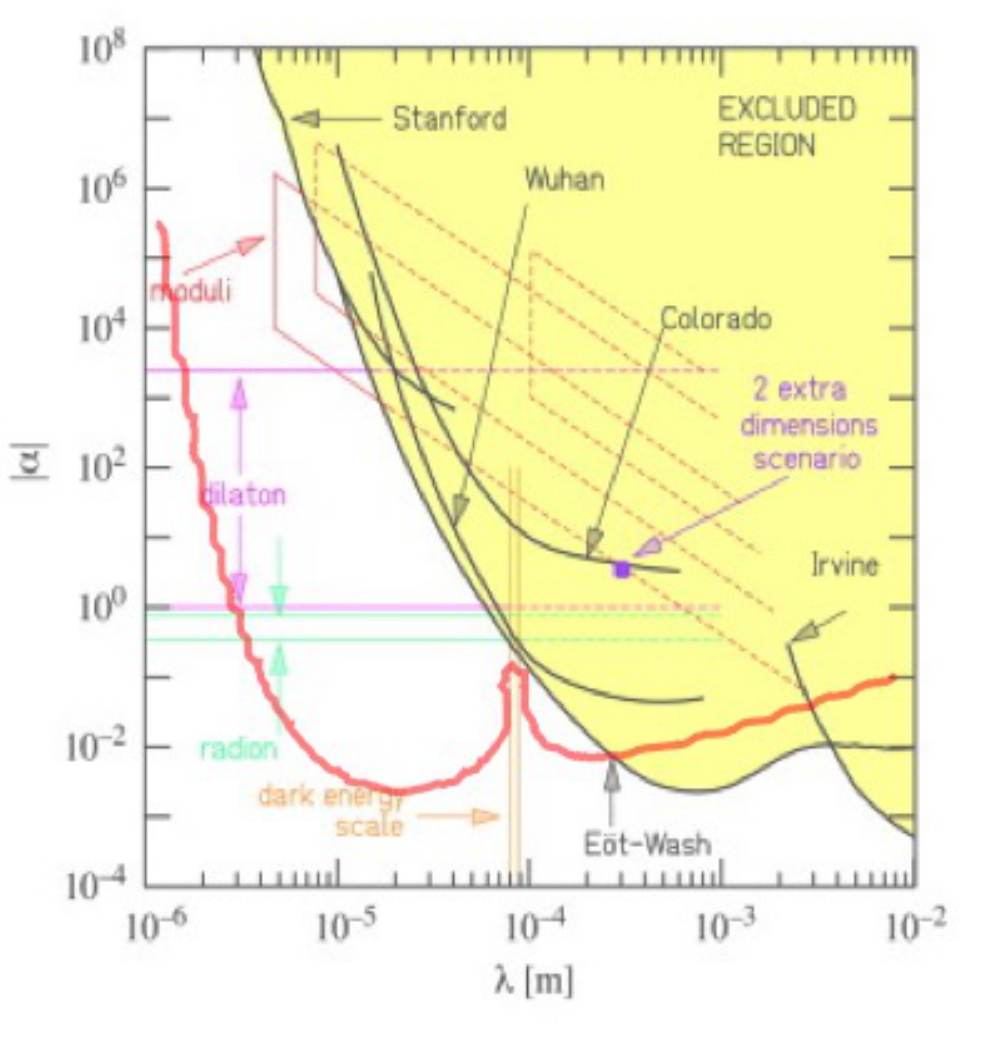} 
\caption{\it  In yellow, we show the 95\% exclusion lines from several experiments on deviations from the 4-dimensional $1/r^2$ Newton's law in the $(\lambda, \alpha)$ plane
(from Ref.~\cite{Adelberger:2009zz}). These lines correspond to experimental bounds from Refs.~\cite{Hoyle:2004,Kapner:2006si,Spero:1980,Hoskins:1985,Tu:2007,Long:2003,Chiaverini:2003,Smullin:2005}. In red we show the 95\% CL bound that can be drawn using the method outlined in this section
for a diamagnetic sphere of mass $m^\prime = 10^{-9}$ g put into orbit around a platinum source of mass $m = 10^{-7}$ g with initial distance $r_0 = 190$ $\mu$m and
initial angular velocity $\dot \theta_0 = 1.8 \times 10^{-4}$ rad/s.
 }
\label{fig:newalphalambdabounds}
\end{figure}

The results obtained in the previous section for the case of gravity in a ${\cal M}_4 \times {\cal S}_1$ space-time with one extra spatial dimension compactified on 
a circle of radius $R$ can be generalized to study any deviation from the $1/r^2$ Newton's law. Consider the case in which two bodies of mass $m$ and $m^\prime$, respectively, are located onto our brane ({\em i.e. here}). In this case, the gravitational potential generated by $m$ and acting on $m^\prime$ is given by eq.~(\ref{eq:PotentialLiu5d}) computed for the special case $y = 0$. When the distance $r$ between the two masses is large compared with the compactification radius ({\em i.e.} $a = r/R \gg 1$), the potential can be approximated with eq.~(\ref{eq:Yukawapotential}). This approximation has 
the same functional form of the Yukawa potential used to parametrize experimentally deviations from the Newton 4-dimensional law:
\begin{equation}
\label{eq:Yukawapotential2}
V_{\rm pheno} (\alpha,\lambda, r) = - \frac{G_4 m \, m^\prime}{r} \left [ 1 + \alpha \, e^{- r/ \lambda} \right ] \, ,
\end{equation}
with the particular choices $\lambda = R$ and $\alpha = 2 \cos d$ ({\em i.e.} $\alpha = 2$ for $d = 0$) and $G_4$ 
related to the fundamental 5-dimensional  coupling by eq.~(\ref{eq:5dimNewtonconstant}). However, eq.~(\ref{eq:Yukawapotential2}) describes 
any model\footnote{Notice that $\alpha$ may be positive or negative.} that introduces small, exponentially suppressed, deviations to the inverse-square Newton's 
law that depend on a single physical scale $\lambda$. The yellow (gray for B\&W printing) )region in Fig.~\ref{fig:newalphalambdabounds} represents bounds at 95\% CL on deviations from the 4-dimensional Newton's law drawn in the $(\lambda, \alpha)$ plane  (taken from Ref.~\cite{Adelberger:2009zz} with bounds obtained in Refs.~\cite{Hoyle:2004,Kapner:2006si,Spero:1980,Hoskins:1985,Tu:2007,Long:2003,Chiaverini:2003,Smullin:2005}). Notice that different theoretical models predict, generically, different expected ranges for $\alpha$. In the particular case of one compact extra spatial dimension, as we have seen, $\alpha = 2$.

In order to apply the results of Sect.~\ref{sec:examples}  and \ref{sec:d0} to study eq.~(\ref{eq:Yukawapotential2}), we sketch the following hypothetical experimental setup: 
\begin{enumerate}
\item Consider a 1 mm$^3$-wide laboratory, with a platinum sphere with radius $r_P = 10.3$ $\mu$m and mass $M_P = 10^{-7}$ g located at the center of the lab;
\item Insert the lab between two magnets, so that we may levitate a diamagnetic satellite in order to cancel the Earth gravitational field\footnote{Possible alternatives 
may be to use an optically-cooled levitating dielectric satellite \cite{Geraci:2010,Yin:2013lqa,Geraci:2014gya}, or to move the mm$^3$-size lab into a zero gravity environment.};
\item Introduce a diamagnetic sphere with mass $m_S = 10^{-9}$ g in the lab so as to match some carefully chosen initial conditions for its distance from 
         the source and its tangential velocity. The diamagnetic sphere can be, for example, made of pyrolitic graphite, with a density $\rho_{\rm PG} = 2.2$ g/cm$^3$
         (for which the radius of the sphere would be $r_S = 4.8$ $\mu$m). In this case, magnets producing a magnetic field $B \sim 0.5$ T suffice to levitate
         the satellite, given the diamagnetic susceptibility of pyrolitic graphite, $\chi = - 16 \times 10^{-5}$  \cite{PhysRev.123.1613,Simon:2000}.
         Introducing the satellite into the lab with given initial conditions is, of course, the most difficult task to achieve experimentally.  
         However, recent results \cite{Kobayashi:2012} show that levitating pyrolitic graphite may be put into motion by means of photo-irradiation.
\end{enumerate}

Once the diamagnetic satellite S is put into motion around the platinum planet P, we connect a trigger to a clock in such a way that every 
time the satellite crosses the line $y = 0$ (at any point on the $x$ axis) the measure of the time needed to S to perform a $2 \pi$-revolution around P is taken.
The error in the measurement of each $T_n$ is the clock sensitivity, neglecting the delay between the trigger and the clock (remember that we are dealing
with revolution times that ranges from minutes to hours). We will consider in the statistical analysis that follows a very conservative $\sigma_T = 1$ s error. 
The collection of $N_{\rm rev}$ revolution times $T_n$ forms our data sample. Once the data are collected, we try to fit our data within the hypothesis that they reproduce
a constant revolution time $T_n = T_N$, being $T_N$ the period of a Newtonian revolution. This is done by computing the following $\chi^2$: 
\begin{equation}
\label{eq:chi2revolution}
\chi^2 = \sum_{n=1}^{N_{\rm rev}} \frac{(T_n - T_N)^2}{1} \, .
\end{equation}
In the following, we have considered $N_{\rm rev} = 20$, that would correspond approximately to a couple of days of data taking in the case of Newtonian orbits.

This procedure can be applied to the Large Extra Dimension case discussed above, but can be also generalized to the case of a phenomenological 
Yukawa potential as the one given in eq.~(\ref{eq:Yukawapotential2}). In this case, the modified gravitational force is: 
\begin{equation}
\label{eq:Yukawaforce}
F_{\rm pheno} (\alpha,\lambda, r) = - \frac{G_4 m \, m^\prime}{r^2} \left [ 1 + \alpha \, \frac{r}{\lambda} \, e^{- r/ \lambda} \right ] \, ,
\end{equation}
where $\alpha = 2 \cos d$ and $\lambda = R$ in the case of a brane-to-brane force, eq.~(\ref{eq:BBforceagg1}).

The results obtained using the setup described above and eq.~(\ref{eq:chi2revolution}) for the initial conditions $r_0 = 190$ $\mu$m 
and $\dot \theta_0 = 1.8 \times 10^{-4}$ rad/s are shown in Fig.~\ref{fig:newalphalambdabounds}. Present bounds, as already said, are represented
by the yellow region, whereas our results at 95\% CL are shown by a red thick line. It can be seen that the bound on 
$\lambda$ can be pushed down to a few microns for any value of $\alpha \geq 10^{-1}$, whereas we get $\lambda < 10$ $\mu$m for $\alpha$ as low as $5 \times 10^{-3}$. 
Below $\lambda = 1$ $\mu$m we lose sensitivity as the exponential factor in the Yukawa potential ${\rm exp} (- r/\lambda)$ rapidly kills the signal
(to go beyond this limit, entering into the nano-world, we should change $m$ and $m^\prime$). For $\lambda > 100$ $\mu$m
there is also a reduction in the sensitivity due to the $r/\lambda$ factor in front of the exponential term in eq.~(\ref{eq:Yukawaforce}). 
On the other hand, for the particular choice of initial conditions
and masses $m_S$ and $m_P$, we have maximal sensitivity for $\lambda$ in the interesting range $\lambda \in [10, 100]$ $\mu$m. Notice that the sensitivity loss 
that can be seen for $\lambda \sim 80$ $\mu$m is due to a cancellation between the Yukawa correction to the gravitational force 
$-(G_N m_S m_P/r^2) \, \alpha \, r/\lambda \, {\rm exp}(-r/\lambda)$ and the centripetal force term $h_0/r^3$ in eq.~(\ref{eq:eombranetobrane})
for the particular choice of the initial conditions. We have eventually checked that our results are independent on the sign of $\alpha$.

An important point to stress is that in eq.~(\ref{eq:chi2revolution}) we have not included backgrounds nor systematic errors. This has not been due to negligence, though.
Even if a more careful study of the possible backgrounds should be performed before implementing the setup proposed here in a real experiment, 
we have thoroughly checked the principal background sources convincing ourselves that they are indeed irrelevant or negligible (for different reasons). 
We list them in order of importance:
\begin{enumerate}
\item First of all, the most important background that limit the sensitivity of experiment searching for deviations from the $1/r^2$ Newton's law is that due to electrostatic forces: these may be Coulombian, dipolar and Van der Waals forces. These forces, for macroscopic objects such those considered in the setup proposed above (our S and P spheres
are indeed much bigger than molecular or atomic scales), have a $1/r^2$ dependence on the distance of S from P. Therefore, for the Bertrand's theorem, they will only
modify the period of the orbit of S around P whilst still maintaining a closed, elliptical orbit with identical times for any revolution of S around P. Deviations from the Newton's
law in the form of a Yukawa potential, on the other hand, will induce a non-elliptical orbit and a precession of the periapsis. A $\chi^2$ analysis
using eq.~(\ref{eq:chi2revolution}), but comparing $T_n$ with the average revolution time $\bar T = 1/N_{\rm rev} \sum_{n=1}^{N_{\rm rev}} T_n$ and not with the Newtonian
period $T_N$, could easily take into account these backgrounds.
\item Another relevant source of background in experiments testing the $1/r^2$ law is the Casimir force acting between the probe and the source of the gravitational field, 
that are usually both conductors. The Casimir force for two conducting spheres has a rather involved dependence on the distance $r$ between the spheres (see, 
for example, Ref.~\cite{Bulgac:2005ku}), that however goes as $1/r^4$ for small distances. This may potentially induce an observable precession of the periapsis.
In our case, however, we use a diamagnetic sphere as the probe, thus reducing significantly any possible Casimir force between the two objects. 
\item Impurities in the magnetic field used to levitate the diamagnetic sphere are randomly distributed along the sphere orbit. Therefore, they should reasonably average out
without affecting the gravitational effects that alter the revolution times $T_n$ pattern. 
\item We have also checked that general relativity effects (similar to those causing the Mercury perihelion precession) are completely negligible in the considered setup.
\end{enumerate}

As a final check, we have parametrized the impact of possible backgrounds in the form of a correction of the Newton force by introducing the following potential: 
\begin{equation}
V = - \frac{G_N m_S m_P}{r} \left [ 1 + \frac{Q_1}{R a}  + \frac{Q_2}{R^2 a^2} + \frac{Q_3}{R^4 a^4} \right ]
\end{equation}
where $Q_1, Q_2$ and $Q_3$ are the (dimensionful) couplings of possible sources of backgrounds in units of the gravitational coupling $G_N m_S m_P$. 
We have found that, in order to have a significant impact on the geometrical and kinematical properties of the orbit, they must be: 
$Q_1/R  > 10^{-1}$,  $Q_2/R^2 > 10^{-1}$ and $Q_3/R^4 > 1$ for $R = 10$ $\mu$m.

In order to realize such an experiment, of course, also systematic errors should be taken into account. This is not the place, however, where to study their
impact on the shown results.

\section{Conclusions}
\label{sec:concl}

This paper, as often occurs, started with a limited goal (to study deviations from Newtonian orbits when dealing with a model in which particles 
are attached to different branes embedded in a compact $(4+n)$-dimensional space-time) to evolve along its completion to something potentially
more ambitious, {\em i.e.} the possibility to detect deviations from the $1/r^2$ Newton's law using precisely the study of departures from Newtonian orbits
in 4-dimensions (regardless of the particular model that may induce these departures). 
In Sects.~\ref{sec:gravipotential} to \ref{sec:examples}, we develop the formalism needed to study 
the kinematical characteristics of orbits for two bodies lying on different branes in a ${\cal M}_4 \times {\cal S}_1$ space-time, with an extra spatial 
dimension compactified on a circle of radius $R$. First, we computed
the gravitational potential in the considered manifold, as it was done in Refs.~\cite{Kehagias:1999my,Floratos:1999bv}. Then, we computed the 
force acting on a mass $m^\prime$ attached to a brane at a distance $d = y/R$ from the source of the gravitational field $m$ located on a brane at $y = 0$. 
This has been done following the outline of Ref.~\cite{Liu:2003rq}. Eventually, in Sect.~\ref{sec:examples} we used these results to study the motion of
a mass $m^\prime \sim 10^{-9}$ g lying onto our brane, orbiting around the projection of a gravitational source $m \sim 10^{-7}$ g
located on a brane at a distance $d = y/R$ from us, with $R = 10$ $\mu$m. The considered masses have been chosen so that Newtonian, elliptical, orbits 
have a typical size ranging from tens to hundreds of microns, {\em i.e.} in a region not yet thoroughly tested experimentally. The compactification radius
is just below the present upper bound on the size of an extra spatial dimension. Even if this setup cannot explain the large hierarchy between the 
electroweak symmetry breaking scale $\Lambda_{\rm ew}$ and the Planck scale $M_P$, the hierarchy problem may still be solved assuming that more the
one extra-dimension exists.
We have found several interesting features: first of all, orbits are not elliptical in a significant portion of the initial conditions parameter space. 
They may be bounded, but are not closed (as guaranteed by the Bertrand's theorem, since correction to the gravitational force have not a $1/r^2$
dependence on the distance). A significant precession of the periapsis (the point at the minimal distance from the source of the gravitational field) is generally observed. 
The distance at the periapsis can be smaller or larger than the corresponding distance in the Newtonian case, depending on the initial conditions. 
In addition to this, the time needed to $m^\prime$ to perform a $2 \pi$-revolution around the projection of $m$ onto our brane is usually 
quite different from the (constant) period find in a Newtonian orbit and it may change from a revolution to the next. 
Therefore, when mock data are produced within a two-brane models
and fitted with a Newtonian model,  we have found that the fit is poor in a significant portion of the parameter space, {\em i.e.} a
Newtonian potential is not able to reproduce the data.This result, of course, depends significantly on the distance between the two branes:  
the nearer, the more difficult the two models are to be distinguished.

Our results seems to imply that the study of the geometrical and kinematical characteristics of orbits in the micro-world may represent a powerful tool to detect
deviations from standard Newtonian dynamics at the micron scale. For this reason, in Sect.~\ref{sec:d0} we have applied the same technique to the interesting
case $d = 0$, {\em i.e.} when both the gravitational source $m$ and the test mass $m^\prime$ lie on the same 4-dimensional manifold embedded in 
a 5-dimensional compact bulk. 
We have found that significant deviations from Newtonian orbits can be observed also in this case, when a reasonable window in the initial 
conditions parameter space is considered. In particular, for particular choices of the initial conditions, extremely large departures from elliptical, stable and 
periodic orbits can be seen. The measurement of the time needed to $m^\prime$ to perform $n$ $2 \pi$-revolutions around $m$ 
gives, therefore, a distinctive, unambiguous signature of modifications of the  $1/r^2$ Newton's law. In order to generalize our results, 
in Sect.~\ref{sec:experiment} we have applied the same technique to the phenomenological Yukawa potential commonly adopted when searching
for departures from the Newton's law. Within this framework, the gravitational potential is modified by an additional term in the form $\alpha G_N m m^\prime \, {\rm exp} (- r/ \lambda)$
where, for the particular case of LED, $\alpha = 2 n$ (being $n$ the number of extra spatial dimensions) and $\lambda = R$. 
Typical bounds on $\lambda$ ranges from $\lambda < 1$ $\mu$m for $\alpha > 10^{10}$ to $\lambda < 100$ $\mu$m for $\alpha \sim 10^{-3}$. In the case $ \alpha = 2$
({\em i.e.} in the case of one LED), we have $\lambda < 44 $ $\mu$m. 
We have therefore proposed a possible experimental setup that could take advantage of the results of the previous sections and that could be used to 
improve our present bounds in the $(\lambda, \alpha)$-plane. The setup consists of a $10^{-7}$ g platinum gravitational source at the centre of 
a 1 mm$^3$ laboratory, inserted between two magnets with a magnetic field $B \sim 0.5$ T so to levitate a $10^{-9}$ g diamagnetic satellite (in order to cancel the Earth gravitational field). 
The satellite is put into orbit around the source at an initial distance $r_0 = 190$ $\mu$m with an angular velocity $\dot \theta_0 = 1.8 \times 10^{-4}$ rad/s 
(where the initial conditions are chosen to maximize the distortion of the orbit with respect to a Newtonian one, whilst avoiding the crash of the satelllte onto the planet surface). 
The resulting orbit is extremely irregular: for $\alpha = 2, \lambda = 10$ $\mu$m, an almost elliptical, very slow, half orbit is followed by a nearly circular, 
very fast, one, such that the revolution times change abruptly from one revolution to the next. The significant gravitational slingshot effect is caused by a 
stronger gravitational force at the periapsis of the orbit. For larger values of $\alpha$ and smaller values of $\lambda$, we have found that 
measuring the first 10 to 20 revolution times seems to be enough to detect small departures from elliptical, periodic orbits and, thus, from the $1/r^2$ Newton's law.
Bounds below a few microns on $\lambda$ can be obtained at 95\% CL for $\alpha > 1$, whereas for $\alpha > 5 \times 10^{-3}$ we can put a limit $\lambda < 10$ $\mu$m
at the same CL (the present bound on $\lambda$ for $\alpha = 10^{-2}$ is $\lambda < 300$ $\mu$m). Although our statistical analysis has been carried out
with no backgrounds, we have checked that the most relevant backgrounds that afflict experiments looking for deviations from the $1/r^2$ Newton's law, 
such as Coulombian, dipolar or Van der Waals forces, Casimir attraction or general relativity corrections, are either irrelevant 
(as they cannot cause a precession of the periapsis or alter the periodicity of the orbit)  or negligible in the considered setup. 

We are therefore convinced that further studies regarding the feasibility of the proposed experiment should be carried on in order to determine
the viability of this technique, that could improve our present bounds on deviations from Newtonian gravity in the micro-world by an order of magnitude or more. 

\section*{Acknowledgements}

We are strongly indebted with A. Cros for discussions regarding some experimental aspects of the paper beyond our expertise. 
We acknowledge also useful discussions with P. Hern\'andez, O. Mena, C. Pe\~na-Garay, N. Rius and M. Sorel. \\
\noindent
 This  work  was  partially  supported  by  grants: \\  
\noindent MINECO/FEDER FPA2012-31686, FPA2014-57816-P, \\ \noindent FPA2015-68541-P, PROMETEOII/2014/050 de la Generalitat Valenciana, 
 MINECO's "Centro de Excelencia Severo Ochoa" Programme under grants SEV-2012-0249 and SEV-2014-0398,
and  the  European  projects H2020-MSCA-ITN-2015//674896-ELUSIVES  and H2020-MSCA-RISE-2015.

\bibliographystyle{h-elsevier}
\bibliography{refbrane2brane}

\end{document}